\documentclass[twocolumn]{aastex631} 

\usepackage{gensymb}
\usepackage{amsmath}
\usepackage{array}
\usepackage{tabularx}
\usepackage{xfrac}
\usepackage{threeparttable}

\begin{document}

\title{Kinematics of HI Envelopes Associated with Molecular Clouds}
\author[0009-0004-7985-6342]{Thummim Mekuria}
\affiliation{Department of Astronomy and Astrophysics \\
University of California, Santa Cruz CA,95064, USA}

\author[0000-0002-6001-598X]{Nia Imara}
\affiliation{Department of Astronomy and Astrophysics \\
University of California, Santa Cruz CA,95064, USA}

\begin{abstract}
We investigate the evolution of molecular clouds through the kinematics of their atomic hydrogen (HI) envelopes, using $^{12}\mathrm{CO}$ and 21-cm emission to trace the molecular and atomic gas, respectively. We measure the large-scale gradients, $\Omega$, in the velocity fields of 22 molecular clouds and their HI envelopes, then calculate their specific angular momenta, $j\propto \Omega R^2$. The molecular clouds have a median velocity gradient of $9.6\times 10^{-2}\ \mathrm{km\ s^{-1}\ pc^{-1}}$, and a typical specific angular momentum of $2.7 \times 10^{24}\ \mathrm{cm^2\ s^{-1}}$. The HI envelopes have smaller velocity gradients than their respective molecular clouds, with an average of $\Omega_\mathrm{HI} = 0.03\ \mathrm{km\ s^{-1}\ pc^{-1}}$, and a median angular momentum of $j_\mathrm{HI} \approx 5.7 \times 10^{24}\ \mathrm{cm^2\ s^{-1}}$. For a majority of the systems, $j_\mathrm{HI} > j_\mathrm{H_2}$, with an average of $j_\mathrm{HI}/j_\mathrm{H_2} = 4$. Their velocity gradient directions tend to be misaligned, indicating that angular momentum is not conserved during molecular cloud formation. Both populations exhibit a $j-R$ scaling consistent with that expected of supersonic turbulence: $j_\mathrm{H_2} \propto R^{1.67\pm 0.22}$, and $j_\mathrm{HI} \propto R^{1.71\pm 0.27}$. Combining our measurements with previous observations, we demonstrate a scaling of $j \propto R^{1.50\pm 0.02}$ in star-forming regions spanning 5 dex in size, $R\in (10^{-3},\ 10^2) \ \mathrm{pc}$. We construct a model of angular momentum transport during molecular cloud formation, and derive the angular momenta of the progenitors to the present-day systems. We calculate a typical angular momentum redistribution timescale of 13 Myr, comparable to the HI envelope free-fall times.
\end{abstract}

\section{Introduction}
\begin{figure*}[ht!]
    \centering
    \includegraphics[width=\linewidth]{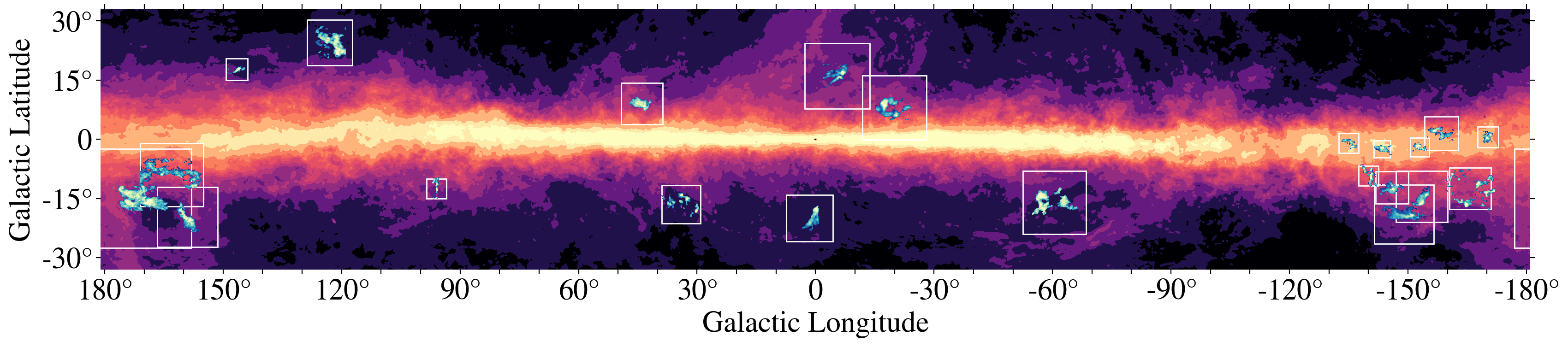}
    \caption{The Solar Neighborhood molecular clouds and their associated HI envelopes studied in this work (Table \ref{tab:intro}). The background map shows the surface density of HI with contours of $\Sigma_\mathrm{HI} \in (5-40)\ M_\odot\ \mathrm{pc^{-2}}$, with an even spacing of $5\ M_\odot\ \mathrm{pc^{-2}}$, as well as at $[50,75,100]\ M_\odot\ \mathrm{pc^{-2}}$. The white squares show the spatial extent of the HI envelope associated with each molecular cloud. (For ease of viewing, molecular gas not analyzed in this work is not included in the map.)}
    \label{fig:mom0}
\end{figure*}
Large-scale linear velocity gradients are routinely observed across molecular clouds and cores (e.g., \citealt{1977ApJ...215..521K, 1993ApJ...406..528G, 2003ApJ...599..258R, 2011ApJ...732...78I, 2011ApJ...732...79I, 2016PASJ...68...24T, 2019ApJ...886..119C, 2020A&A...633A..17B}). These gradients are often attributed to solid-body rotation, and used to infer the specific angular momentum, $j \propto \Omega R^2$, of these structures (e.g., \citealt{1993ApJ...406..528G,2016PASJ...68...24T}). 
The angular momentum of molecular clouds is of interest because it may have a significant impact on the outcomes of star-formation---e.g., the degree of fragmentation \citep{2008ApJ...677..327M}, the IMF \citep{2025ApJ...988..125S}, and proto-planetary disks sizes \citep{2024ApJ...972L..27Y}. 

Observations and theoretical studies agree that molecular clouds form when atomic hydrogen (HI) accumulates to surface densities exceeding $10\ M_\odot \ \mathrm{pc^{-2}}$, sufficient to shield the $\mathrm{H_2}$ molecule from dissociation by interstellar radiation \citep{2009ApJ...693..216K,2016ApJ...829..102I,2021ApJ...920...83S,2023ApJ...955...59P}. \citet[hereafter IB]{2011ApJ...732...78I} and \citet[hereafter IBB]{2011ApJ...732...79I} were the first to compare the kinematics of the molecular and atomic phases of hydrogen involved in the formation of molecular clouds. To do so, they devised a method for identifying the HI associated with molecular clouds using spatial and kinematic criteria. IB studied five star forming regions in the Solar Neighborhood, finding that the HI envelopes had an average velocity gradient of $0.04\ \mathrm{km\ s^{-1}\ pc^{-1}}$. The specific angular momenta they measured for the HI envelopes, $j_\mathrm{HI} \approx 1.9\times 10^{25}\ \mathrm{cm^2\ s^{-1}}$, were $2-5$ times larger than the specific angular momenta of corresponding molecular clouds. 

IBB identified the HI envelopes associated with 45 molecular clouds in M33 \citep{2003ApJ...599..258R}, and measured an average velocity gradient of $\Omega_\mathrm{HI} \approx 0.05\ \mathrm{km\ s^{-1}\ pc^{-1}}$. They found that the HI envelopes had $j_\mathrm{HI} \approx 7.8\times 10^{25}\ \mathrm{cm^{2}\ s^{-1}}$, which is on average 27 times larger than the specific angular momentum of their corresponding molecular clouds. Both IB and IBB also found that the rotation axis of a molecular clouds and its HI envelope within a given system tend to be randomly oriented with respect to one another. 

Had the molecular clouds formed by simple top-down collapse of atomic gas, the angular momentum of the HI would have been conserved in the molecular clouds. However, IB and IBB demonstrated that neither the magnitudes nor the orientation of the angular momenta of the atomic and molecular gas are aligned. This so-called ``angular momentum problem” is a long-standing puzzle \citep{1981MNRAS.194..809L,1999A&AS..134..241P,2007ApJ...654..240R,2011ApJ...732...79I}, and has been attributed to gravitational (e.g., \citealt{1984MNRAS.206..197L}), magnetic (e.g., \citealt{1979ApJ...230..204M,2024ApJ...963..106M}), and turbulent (e.g., \citealt{2024arXiv240810406V}) torques braking molecular clouds during formation.

Furthermore, IBB measured the specific angular momenta of isolated HI clouds and found that they had lower rotational energies compared to their counterparts associated with molecular emission. The increased shear and rotational energy in the HI envelopes harboring molecular clouds suggests that the molecular clouds play a role in increasing the angular momentum of the HI in their vicinity. Magnetohydrodynamic simulations corroborate this view,  demonstrating that molecular clouds dissipate angular momentum as they evolve \citep{2022ApJ...925...78A,2024ApJ...963..106M}.

In this work, we investigate the kinematics of molecular clouds in the Solar Neighborhood to determine whether the findings presented by IB persist with higher resolution and higher sensitivity HI observations---which allow us to capture a larger dynamic range in density---and a sample of clouds that is $\geq 4$ times larger. We go further than IB, exploring how specific angular momentum varies in star-forming regions over a wide range of evolutionary stages, and estimating the timescale for angular momentum redistribution during molecular cloud formation.

The paper is organized as follows: in Section \ref{sec:data}, we present the catalog of molecular clouds we study and the observational data used to trace molecular and atomic hydrogen. Section \ref{sec:HIs} describes the spatial and kinematic criteria used to select the HI envelopes, as well as how we measure the physical properties of each population. Velocity field maps and kinematic properties of both populations are presented in Section \ref{sec:kinematics}. We quantify population level trends in Section \ref{sec:j_trends}, including a meta-analysis of angular momentum measurements in the literature. In Section \ref{sec:discussion} we develop a physical model to follow angular momentum transport in the molecular and atomic phases, then contextualize our measurements within the large-scale motion of the Galaxy. We end with a summary in Section \ref{sec:summary}.


\section{Data}\label{sec:data}
\subsection{Molecular cloud sample}
We select molecular clouds from the \cite{2019ApJ...879..125Z} catalog of dust-derived distances to Solar Neighborhood molecular clouds. Of the 27 regions targeted in \cite{2019ApJ...879..125Z}, we exclude those located at extreme latitudes ($|b|>30\degree$) or along lines of sight where the $^{12}\mathrm{CO}$ emission blends with the Galactic plane. The final sample (Table \ref{tab:intro}, Figure \ref{fig:mom0}) consists of 22 molecular clouds, including the five in \cite{2011ApJ...732...78I}, as well as other well-studied clouds such as $\rho$-Ophiuchus, the Polaris Flare, and Corona Australis.

\subsection{$^{12}\mathrm{CO}$ observations} 
The Dame, Hartmann, Thaddeus survey \citep[hereafter DHT]{2001ApJ...547..792D} is the most complete $^{12}\mathrm{CO} \ (J = 1 \rightarrow 0)$ map of the Galaxy, capturing emission at latitudes $\leq 30 \degree$ with an angular resolution of $\theta = 8'.5$. This corresponds to a linear resolution of 0.3 - 4 pc for the cloud distances in our catalog (150 - 2000 pc). The velocity resolution of the DHT is $\mathrm{d}v = 1.3\ \mathrm{km s^{-1}}$, and the data cover emission within $|v| \leq\ 300\ \mathrm{km\ s^{-1}}$ of the 115 GHz line center. Analyses are performed on the moment masked datacubes \footnote{DHT is available as position-position-velocity (ppv) datacubes at https://lweb.cfa.harvard.edu/rtdc/CO}  \citep{2011arXiv1101.1499D}. 
\subsection{21-cm observations}
HI4PI \citep{2016A&A...594A.116H} is the highest-sensitivity ($\sigma_\mathrm{rms} = 0.43 \ \mathrm{mK}$) and highest-resolution ($\theta = 16'.2$)  21-cm map of the sky available to date. HI4PI outperforms the LAB survey \citep{2005A&A...440..775K}, which IB used, by a factor of 8 in resolution. HI4PI covers emission within $|v| \leq\ 600\ \mathrm{km\ s^{-1}}$, at a similar spectral resolution as DHT ($1.29\ \mathrm{km\ s^{-1}}$). We also resample the data onto a $0.^{\degree}125$ grid to match the spatial resolution of DHT.\footnote{Measurements made using the resampled data are within a few percent of those made using the native resolution data.} HI4PI is available as fits and HEALPix cubes in various projections \citep{https://doi.org/10.26093/cds/vizier.35940116}; we use the fits cube in galactocentric coordinates, in cartesian projection for all subsequent analysis.

\begin{table*}[ht!]
\begin{center}
\caption{Location of molecular clouds and velocities of HI envelopes}
\begin{tabular}{|l|cc|cc|c|cc|}
\hline
 Name&  $l_0$&$l_1$&  $b_0$&$b_1$&  $d$&$\overline{v}_\mathrm{HI}$& $\sigma_{v,\mathrm{HI}}$\\ 
\hline
 &  \multicolumn{2}{c|}{$\mathrm{deg}$}&  \multicolumn{2}{c|}{$\mathrm{deg}$}&   $\mathrm{pc}$&\multicolumn{2}{c|}{$\mathrm{km\ s^{-1}}$}\\ 
 \hline
 \hline
 Aquila South&  41&29&  -20&-12&  $133\pm 7.3$&1.9& 2.5\\ 
 California& 171&155&  -13&-2&  $470\pm 24.1$&-3.3& 12.2\\ 
 Camelopardalis& 150&144&  16&19&  $213\pm 13.6$&-0.5& 3.3\\ 
 Canis Major OB1&-132.5&-138.5&  -2.5&1&  $1209\pm60.1$&15.9& 6.1\\ 
 Chamaeleon& -54&-67&  -20&-12&  $183\pm9.5$&2.0& 3.3\\ 
 Corona Australis&  4&-1& -24&-16&  $151\pm8.5$&6.0& 3.7\\ 
 Crossbones&-138&-142& -12&-6.5&  $886\pm44.2 $&11.1& 6.4\\ 
 Gemini OB1&-164&-174& -3&3&  $1786\pm89.1$&7.6& 10.4\\ 
 Hercules&  49&41& 7&11&  $227\pm11.0$&5.3& 5.1\\ 
 Lacerta&  98&94& -18&-8&  $503\pm25.5$&0.1& 5.3\\ 
 Lupus& -14&-27& 5&13&  $189\pm12.7$&3.4& 5.1\\ 
 Maddalena& -140.5& -146.5& -5& -0.5&  $2072\pm104.2$&24.7& 11.4\\
 Monoceros OB1& -155& -162& -0.5& 3.5&  $745\pm37.1$&8.8& 7.3\\
 Monoceros R2& -141& -149& -14.5& -10&  $778\pm39.1$&8.7& 6.1\\
 Ophiuchus& 0& -9& 12& 19&  $144\pm7.3$&2.8& 5.1\\
 Orion A& -141& -155& -21& -16&  $432\pm22.1$&7.1& 6.3\\
 Orion B& -149& -158& -17& -10&  $423\pm21.1$&8.3& 5.8\\
 $\lambda$-Orionis& -159.5& -173& -18& -7&  $402\pm20.1$&6.6& 7.3\\
 Perseus& 162& 155& -26& -16&  $294\pm15.1$&3.6& 5.0\\
 Polaris& 127& 118& 20& 30&  $352\pm18.0$&-3.6& 4.9\\
 Rosette& -151& -155& -3.5& -0.5&  $1304\pm65.2$&13.2& 10.2\\
 Taurus& 177& 164& -20& -11&  $141\pm7.3$&5.0& 6.1\\
 \hline
\end{tabular}
\begin{tablenotes}
    Longitude ($l)$ and latitude ($b$) and distances of the molecular clouds studied here, adapted from \cite{2019ApJ...879..125Z}. The line-center ($\overline{v_\mathrm{HI}}$) and linewidth $\overline{\sigma_\mathrm{HI}}$ are calculated by fitting a plane to the velocity spectrum of the HI envelope associated with each molecular cloud.
\end{tablenotes}

\label{tab:intro}
\end{center}
\end{table*}


\section{Methodology}\label{sec:HIs}

Here we describe the spatial and kinematic criteria used to select the molecular clouds and their associated HI. We then present the physical properties of both populations and compare with previous findings.

\subsection{Identifying molecular clouds}
The latitude and longitude ranges that we use to define the boundaries of the molecular clouds (Table \ref{tab:intro}) are guided by those given in \cite{2019ApJ...879..125Z}. Given that $^{12}\mathrm{CO}$ becomes optically thick in molecular clouds at $\mathrm{H_2}$ column densities of $\sim 10^{21}\ \mathrm{cm^{-2}}$ \citep{2016ApJ...829..102I}, we use this threshold to define the lowest-level contour of the molecular clouds in the $^{12}\mathrm{CO}$ observations. For the majority of the clouds in our catalog, we mask data below $N_\mathrm{H_2}= 10^{21}\ \mathrm{cm^{-2}}$. However, there are 9 smaller diffuse molecular clouds in our sample (Aquila South, Camelopardalis, Chamaeleon, Corona Australis, Crossbones, Hercules, Lacerta, $\lambda$-Orions, Polaris) with a significant fraction of their surface area containing emission below $10^{21}\ \mathrm{cm^{-2}}$; for these we use a lower threshold of $\log (N_\mathrm{H_2}) = 20.5$. 

\subsection{Spatial and kinematic bounds of HI envelopes}\label{sec:rhi_vs}
Observations (e.g., \citealt{2016ApJ...829..102I,2023ApJ...955...59P}) show that molecular clouds in the Milky Way tend to be spatially and kinematically co-located with thick envelopes of atomic gas having surface densities of $\geq 10\ M_\odot \ \mathrm{pc^{-2}}$. For a given molecular cloud, IB and IBB define its HI envelope as the atomic gas located inside a cylindrical volume, ``the accumulation region," centered on the cloud. The cylinder has a radius of $R_\mathrm{HI}$, a height of $2R_\mathrm{HI}$, and is oriented with its circular faces parallel to the midplane. Given that the scale height of HI is larger than $\mathrm{H_2}$, the HI in the accumulation cylinder is expected to collapse along the z-axis when contracting \citep{1993prpl.conf..125B,2003ApJ...599..258R,2011ApJ...732...78I}. The accumulation radius, $R_\mathrm{HI}$, is set such that the mass of atomic hydrogen enclosed within the volume, $M_\mathrm{HI}$, is at minimum equal to the molecular cloud mass, $M_\mathrm{H_2}$. 

We adopt the above definition and calculate the radius of an HI envelope as:

\begin{align}\label{eq:R_HI}  
    R_{\mathrm{HI}} = \left( \frac{M_\mathrm{H_2}}{2\pi n_\mathrm{_{HI}} \mu m_\mathrm{p}} \right)^{1/3}, 
\end{align}  

where $n_\mathrm{HI}$ is the number density of atomic hydrogen, $\mu$ is the mean molecular weight, $m_\mathrm{p}$ is the mass of a hydrogen atom, and $M_\mathrm{H_2}$ is the molecular cloud mass. We adopt a hydrogen number density of $n_\mathrm{HI} = 0.57\ \mathrm{cm^{-3}}$, a typical value for the warm neutral medium \citep{2003ApJ...587..278W}. The mean molecular weight is set for Solar metallicity: $\mu = 1.36$.

\begin{figure}[ht!]
\includegraphics[width=\linewidth]{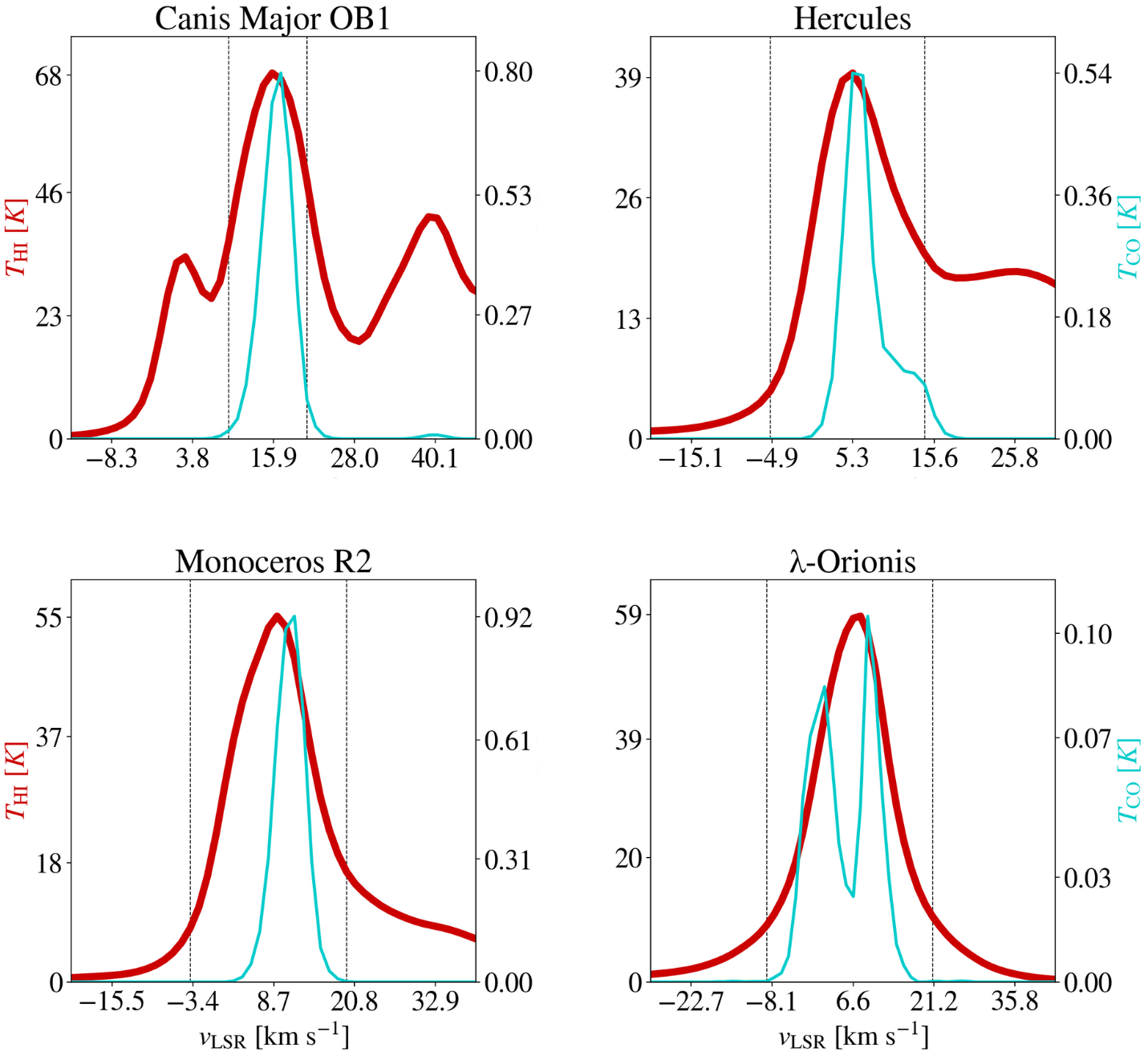}
\caption{Velocity spectra of select molecular clouds (blue) and corresponding HI envelopes (red). The vertical lines indicate velocity bounds used to define the HI envelope. Spectra of the remaining molecular clouds and HI envelopes can be found in Appendix \ref{sec:appndx_specs}.}
\label{fig:specs}
\end{figure}

On sky, the accumulation region will appear as a square with a side length of $2R_\mathrm{HI}$. To identify the atomic gas within this aperture that is kinematically associated with a given molecular cloud, we start by examining the velocity spectrum (Figure \ref{fig:specs}). We find that in all the systems, the emission peaks of the molecular cloud and its HI envelope spectra overlap with one another. 

By fitting a Gaussian profile to each HI spectrum, we derive its line-center, $\overline{v_\mathrm{HI}}$, and linewidth, $\sigma_\mathrm{HI}$. We then define emission within $\overline{v_\mathrm{HI}}\pm 2\sigma_{\mathrm{HI}}$ to be kinematically associated with the molecular cloud. The 21-cm spectra in the direction of some distant clouds (Canis Major OB1, Maddalena, Monoceros OB1, Rosette) consist of multiple emission peaks, likely due to blending with foreground HI. For these clouds, we constrain our selection of kinematically associated HI gas to velocity channels within $\pm 1\sigma_\mathrm{HI}$ of the 21-cm peak that lines up with the $^{12}\mathrm{CO}$ spectrum of the corresponding molecular cloud.

\subsection{Mass and radius}
The mass of a molecular or atomic cloud is calculated as

\begin{align}\label{eq:M}
M &= \mu m_\mathrm{p} \sum^{N_\mathrm{pix}} N_\mathrm{H} \ \mathrm{d} A,
\end{align}

where $N_\mathrm{H}$ is the column density of hydrogen per pixel, and $\mathrm{d}A$ is the pixel area. The latter is equal to $(d\ \mathrm{d} s)^2$, where $d$ is the distance and $\mathrm{d} s$ is the pixel size. For a molecular cloud, the effective radius is determined from the total $^{12}\mathrm{CO}$ emitting area: $R_\mathrm{H_2} = \sqrt{(N_\mathrm{pix}\ \mathrm{d} A)/\pi}$.

The molecular clouds column density, $N_\mathrm{H_2}$, is derived from the zeroth moment of the $^{12}\mathrm{CO}$ data:

\begin{align}\label{eq:N}
N_\mathrm{H_2} = X_\mathrm{CO} \sum T_\mathrm{B}(v)\  \mathrm{d} v,
\end{align}

where $X_\mathrm{CO}$ is the metallicity-dependent CO-to-$\mathrm{H_2}$ conversion factor, $T_\mathrm{B}$ is the brightness temperature, and $\mathrm{d} v$ is the velocity resolution. We adopt a value of $X_\mathrm{CO} =1.97 \times 10^{20}\ \mathrm{cm^{-2}\ (K\ km\ s^{-1})^{-1}}$, typical for solar metallicity gas \citep{2022ApJ...931....9L}. The HI column density is calculated from the 21-cm emission, assuming the optically thin limit \citep{2018MNRAS.480L.126S}: 

\begin{align}
    N_\mathrm{HI} = 1.82 \times 10^{18} \sum T_\mathrm{B}(v)\  \mathrm{d} v 
\end{align}

We list the masses and sizes of the molecular clouds and their atomic envelopes in Table 2. The molecular clouds have an average size of $17$ pc, with the population spanning $R_\mathrm{H_2}\in (4,\ 43)\ \mathrm{pc}$. The HI accumulation radii are on average $2.8\times$ larger, with a mean of $41$ pc, and $R_\mathrm{HI}$ lying in the range $(11,\ 83)\ \mathrm{pc}$. The molecular clouds cover a large range of masses: $M_\mathrm{H_2}\in (580,\ 2.7\times 10^5)\ M_\odot$, with an average of $5.8 \times 10^4\ M_\odot$ and a median of $2.8\times 10^4\ M_\odot$. The HI envelopes occupy a smaller range, $M_\mathrm{HI} \in (2.3\times 10^3,\ 1.4\times 10^6)\ M_\odot$, with a mean of $1.8\times 10^5\ M_\odot$ and median of $8.1\times 10^4\ M_\odot$. 

On average, the HI masses measured from 21-cm emission are $8$ times larger than the corresponding $M_\mathrm{H_2}$. The latter was used as a lower limit to calculate the HI accumulation radius (Equation \ref{eq:R_HI}). A potential reason for this discrepancy could be that we underestimated the atomic hydrogen density when calculating $R_\mathrm{HI}$. One could use the denser component of the neutral ISM, with $n_\mathrm{CNM} \sim 40\times  n_\mathrm{WNM}$ \citep{2003ApJ...587..278W}, to trace the HI more directly associated with a molecular cloud. Since CNM is best traced in absorption (e.g., \citealt{2024ApJ...973L..27S,2014ApJ...793..132S}), surveys of HI self-absorption with comparable coverage and resolution to the $^{12}\mathrm{CO}$ observations would be useful for such an analysis. A different reason could be that unrelated 21-cm emission along the line-of-sight is contributing to the integrated column density $N_\mathrm{HI}$ and thereby increasing the calculated HI envelope mass. Line-of-sight distance maps with comparable coverage and resolution to the $^{12}\mathrm{CO}$ observations may be used to distinguish between the HI in the vicinity of the molecular cloud and unassociated atomic gas. 
\begin{figure*}[ht!]
    \centering
    \includegraphics[width=\linewidth]{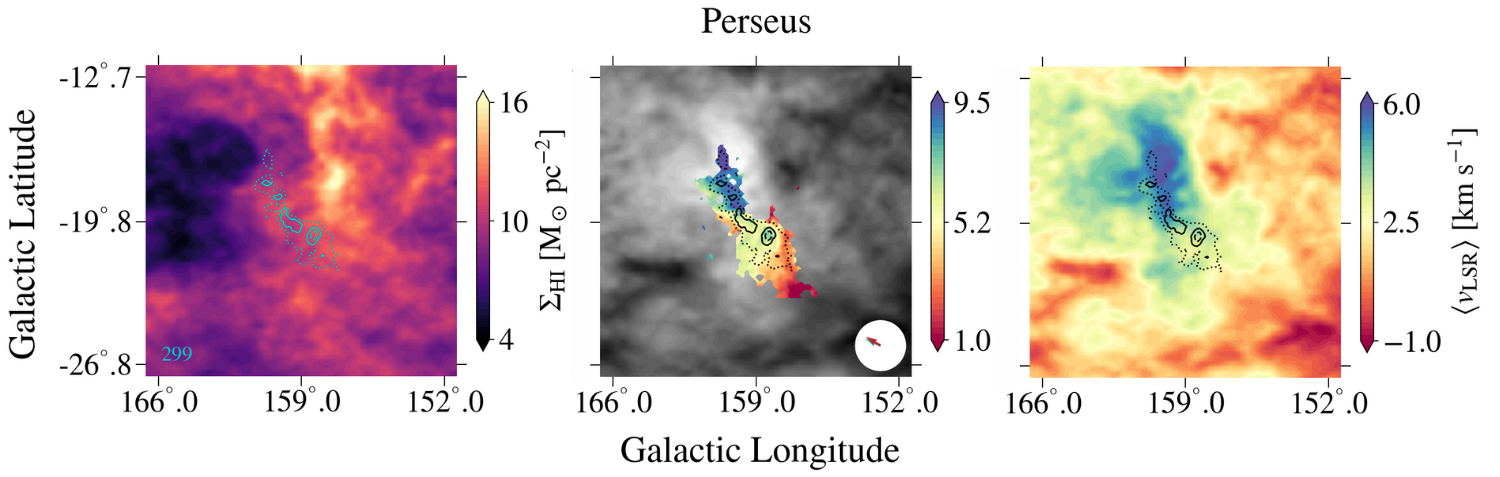}
    \caption{\textit{Left}: Column density map of the HI envelope around the Perseus molecular cloud (black outline). The overlaid blue contours are at $(0.25,\ 0.5,\ 0.75)$ times the peak $\mathrm{H_2}$ surface density, which is indicated in lower left corner. \textit{Middle}: Velocity field maps of the Perseus molecular cloud and (\textit{right}) HI envelope, derived from the first moments of $^{12}\mathrm{CO}$ and 21-cm emission at each pixel (Equation \ref{eq:first_mom}). The blue (red) arrow in the lower right corner of the middle panel indicates the velocity gradient direction $\theta_\mathrm{H_2}$ ($\theta_\mathrm{HI}$). Maps of the remaining molecular clouds and HI envelopes can be found in Appendix \ref{sec:appndx_maps}.}
    \label{fig:maps}
\end{figure*}

\begin{table*}\label{tab:m_r_omg_j}
\centering
 \caption{Physical Properties of Molecular Clouds and their HI envelopes.}

\begin{tabular}{|l|cc|cc|cc|cc|cc|cc|} 

\hline
 Name&  $M_\mathrm{H_2}$&  $M_\mathrm{HI}$&$R_\mathrm{H_2}$&  $R_\mathrm{HI}$&$\Omega_\mathrm{H_2}$ & $\Omega_\mathrm{HI}$&$\theta_\mathrm{H_2}$ &  $\theta_\mathrm{HI}$  &  $\rho_\mathrm{H_2}$&$\rho_\mathrm{HI}$ & $j_\mathrm{H_2}$&$j_\mathrm{HI}$ \\
 \hline
 &  \multicolumn{2}{c|}{$10^4\ M_\odot$}&\multicolumn{2}{c|}{$\mathrm{pc}$ }&\multicolumn{2}{c|}{$10^{-2}\  \mathrm{km\ s^{-1}\ pc^{-1}}$}&\multicolumn{2}{c|}{$\mathrm{deg}$ } & \multicolumn{2}{c|}{} & \multicolumn{2}{c|}{$10^{24}\ \mathrm{cm^2\ s^{-1}}$ }\\ 
 \hline
 \hline
 AqS&  0.07&   0.23&$4.4\pm0.2$&   $11.5\pm0.4$&$7.9 \pm 0.6$& $7.8\pm 0.05$&108.0&  118.0& 0.76&0.99 & $0.19\pm0.02$&$1.28\pm0.09$\\ 
 Cal&  14.37&   41.19&$32.0\pm1.6$&   $66.8\pm2.3$&$10.2 \pm 0.1$& $10.7\pm0.02$&-6.4&  -74.1& 0.99&0.99 & $12.92\pm1.33$&$58.91\pm4.03$\\ 
 Cam&  0.06&   0.24&$3.6\pm0.2$&   $10.7\pm0.5$&$10.9 \pm 2.4$& $3.2\pm0.07$&-103.2&  -80.2& 0.53&0.77 & $0.18\pm0.05$&$0.45\pm0.04$\\ 
 CMO&  7.97&   14.83&$23.0\pm1.1$&   $54.9\pm1.8$&$3.3 \pm 0.5$& $1.7\pm0.02$&-63.0&  73.8& 0.27&0.91 & $2.19\pm0.38$&$6.47\pm0.44$\\ 
 Chm&  0.82&   1.56&$9.2\pm0.5$&   $25.7\pm0.9$&$10.5 \pm 0.2$& $2.8\pm0.02$&162.7&  166.5& 0.6&0.78 & $1.08\pm0.11$&$2.30\pm0.16$\\ 
 CrA& 0.19&  0.54&$5.0\pm0.3$&   $15.8\pm0.6$&$8.8 \pm 0.6$& $1.1\pm0.04$&135.8&  -166.4& 0.62&0.4 & $0.27\pm0.03$&$0.33\pm0.03$\\ 
 Cbn& 3.11&  12.26&$22.6\pm1.1$&   $40.1\pm1.3$&$5.8 \pm 0.2$& $5.4\pm0.05$&21.0&  100.1& 0.96&0.97 & $3.64\pm0.38$&$10.62\pm0.71$\\ 
 GmO& 27.46&  135.5&$38.3\pm1.9$&   $82.9\pm2.8$&$6.0 \pm 0.3$& $3.2\pm0.02$&-38.3&  -27.4& 0.74&1.0 & $10.79\pm1.22$&$27.52\pm1.84$\\ 
 Her& 0.46&  1.68&$7.2\pm0.3$&   $21.2\pm0.7$&$33.3 \pm 1.3$& $2.9\pm0.04$&-158.6&  -90.3& 0.66&0.81 & $2.12\pm0.22$&$1.62\pm0.11$\\ 
 Lac& 0.56&  2.03&$9.7\pm0.5$&   $22.7\pm0.8$&$23.0 \pm 1.4$& $5.6\pm0.08$&26.1&  106.7& 0.93&0.97 & $2.64\pm0.31$&$3.53\pm0.24$\\ 
 Lup& 0.98&  4.06&$8.7\pm0.6$&   $27.3\pm1.2$&$16.2 \pm 0.4$& $6.8\pm0.03$&97.3&  80.8& 0.82&0.81 & $1.53\pm0.21$&$6.27\pm0.56$\\ 
 Mad& 22.0&  58.5&$43.0\pm2.2$&   $77.0\pm2.6$&$10.7 \pm 0.6$& $2.7\pm0.03$&75.2&  95.4& 0.74&0.95 & $24.45\pm2.82$&$20.03\pm1.36$\\ 
 MOB& 8.35&  16.58&$21.1\pm1.1$&   $55.8\pm1.9$&$12.1\pm 0.8$& $0.5\pm0.02$&92.6&  -108.8& 0.48&0.55 & $6.64\pm0.79$&$2.01\pm0.15$\\ 
 MR2& 9.11&  19.89&$23.3\pm1.2$&   $57.4\pm1.9$&$6.4 \pm 0.2$& $3.1\pm0.02$&-153.5&  97.3& 0.93&0.94 & $4.29\pm0.44$&$12.78\pm0.86$\\ 
 Oph& 0.44&  1.94&$5.3\pm0.3$&   $21\pm0.7$&$30.5 \pm 0.5$& $5.7\pm0.03$&-128.1&  -70.1& 0.76&0.96 & $1.06\pm0.11$&$3.1\pm0.21$\\ 
 OrA& 9.01&  16.34&$20.1\pm1.0$&   $57.2\pm2.0$&$8.9\pm 0.2$& $1.6\pm0.02$&-166.0&  136.7& 0.62&0.84 & $4.44\pm0.47$&$6.49\pm0.45$\\ 
 OrB& 5.42&  14.62&$13.2\pm0.7$&   $48.3\pm1.6$&$0.9\pm 0.4$& $2.2\pm0.03$&111.6&  85.2& 0.12&0.75 & $0.20\pm0.09$&$6.24\pm0.42$\\ 
 $\lambda$Or& 2.55&  10.51&$18.5\pm0.9$&   $37.6\pm1.2$&$20.6\pm 0.3$& $2.9\pm0.02$&83.6&  -0.9& 0.64&0.96 & $8.68\pm0.87$&$5.05\pm0.34$\\ 
 Per& 3.03&  5.72&$11.5\pm0.6$&   $39.8\pm1.4$&$23.9\pm 0.5$& $3.7\pm0.03$&24.2&  31.1& 0.96&0.93 & $3.92\pm0.41$&$7.23\pm0.50$\\ 
 Pol& 2.2&  2.3&$18.6\pm1.0$&   $35.7\pm31.2$&$6.7\pm 0.2$& $2.4\pm0.02$&-4.2&  22.0& 0.76&0.91 & $2.85\pm0.30$&$3.84\pm0.26$\\ 
 Ros& 8.86&  3.0.23&$24.3\pm1.2$&   $56.9\pm1.9$&$8.7\pm 0.5$& $1.3\pm0.02$&-150.5&  99.5& 0.53&0.93 & $6.31\pm0.74$&$5.36\pm0.37$\\ 
 Tau& 1.49&  5.34&$9.7\pm0.5$&   $31.4\pm1.1$&$8.4\pm 0.2$& $8.2\pm0.03$&-119.2&  -62.8& 0.77&0.95 & $0.97\pm0.1$&$9.90\pm0.68$\\ 
\hline
    \end{tabular}
    \begin{tablenotes}
    \item Notes: Physical properties calculated from $^{12}\mathrm{CO}$ and 21-cm observations. The properties listed include cloud mass ($M$), radius ($R$), velocity gradient magnitude ($\Omega$) and direction ($\theta$), Pearson coefficient ($\rho$), and specific angular momentum ($j$). The uncertainties in $R$ are calculated from the distance uncertainties reported in \citep{2019ApJ...879..125Z}, and propagated to $j$. The latter is also dependent on the errors of the fit to Equation \ref{eq:v_plane}, which propagate through $\Omega$.
    \end{tablenotes}
\end{table*}


\section{Velocity Gradients}\label{sec:kinematics}

Both molecular clouds and their associated HI envelopes are observed to have systematic, near-linear gradients in the velocity fields, often covering the whole extent of the cloud (e.g., \citealt{1977ApJ...215..521K,1993ApJ...406..528G,2011ApJ...732...79I,2018A&A...612A..51B}). In the following, we derive the velocity fields and calculate large-scale gradients of the molecular clouds and their atomic envelopes. 

The velocity gradient $\overrightarrow{\Omega}$ is measured from systematic variations in the first moment map, which maps the intensity-weighted centroid of emission at each pixel:

\begin{align}\label{eq:first_mom}
\langle v_\mathrm{LSR} \rangle = \frac{\sum v\ T_\mathrm{B}(v)\ \mathrm{d} v}{\sum T_\mathrm{B}(v)\ \mathrm{d} v}, 
\end{align}

We follow \cite{1993ApJ...406..528G} and calculate $\Omega$ by fitting a plane, given below, to the first moment maps.

\begin{align}\label{eq:v_plane}
v(x,y) = v_0 + \Omega_x (x-x_0) + \Omega_y (y-y_0),
\end{align}

Above, $(x_0, y_0)$ is the center-of-mass (COM) of the cloud, $v_0$ is velocity of the COM, and $\Omega_{x}$ and $\Omega_y$ are the gradient of $\langle v_\mathrm{LSR} \rangle$ along Galactic longitude and latitude, respectively. $\Omega_x, \Omega_y$, and $v_0$ and their respective uncertainties are obtained from a least-squares fit of Equation \ref{eq:v_plane}. The total magnitude and direction of the velocity gradient are calculated from these parameters as:

\begin{align}\label{eq:omg_tht}
    |\Omega| &= \sqrt{\Omega_x^2 + \Omega_y^2}\\
    \theta &= \tan^{-1} \left( \sfrac{\Omega_y}{\Omega_x} \right).
\end{align}

$\theta $ is oriented such that the rotation axis of the cloud lies along $\phi \equiv (\theta - 90\degree$). In the rest of the text, we use the $\overrightarrow{\Omega}$ to refer to the velocity gradient direction in vector form, and $\Omega$ when referring to the magnitude.

\subsection{Results of gradient fitting}
The magnitude and directions of the velocity gradients are listed in of Table 2. The molecular clouds have gradients in the range $(0.09,\ 0.3)\ \mathrm{km\ s^{-1}\ pc^{-1}}$, with an average of $\Omega_\mathrm{H_2} = 0.12\ \mathrm{km\ s^{-1}\ pc^{-1}}$ and a median of $0.96\ \mathrm{km\ s^{-1}\ pc^{-1}}$. The HI envelopes have smaller velocity gradients than their molecular counterparts, $\Omega_\mathrm{HI}\  \in (0.005,\ 0.11)\ \mathrm{km\ s^{-1}\ pc^{-1}}$, with a typical value of $0.03\ \mathrm{km\ s^{-1}\ pc^{-1}}$. We do not find any correlation between $\Omega_\mathrm{H_2}$ and  $\Omega_\mathrm{HI}$.

If these velocity gradients arise due to solid body rotation, the typical rotational period of the molecular clouds is $2\pi/\Omega_\mathrm{H_2} \approx 65\ \mathrm{Myr}$; an order of magnitude larger than both their median sound crossing time $(R/\sigma)_\mathrm{H_2} \approx 5.7\ \mathrm{Myr}$ and median free-fall time $t_\mathrm{ff,H_2} \approx 5.6\ \mathrm{Myr}$. The HI envelopes have similar crossing times to the molecular clouds, $(R/\sigma)_\mathrm{H_2} \approx 5.8\ \mathrm{Myr}$ and longer free-fall times, $t_\mathrm{ff,HI} \ \approx 13.8\ \mathrm{Myr}$. Due their smaller gradients, the HI envelopes have longer periods, $2\pi/\Omega_\mathrm{HI} \approx 200\ \mathrm{Myr}$. In both populations, large-scale rotation is sub-dominant compared to turbulent and in-fall motions, although the discrepancy is decreased in the molecular clouds.

\begin{figure}[ht!]
    \centering
    \includegraphics[width=\linewidth]{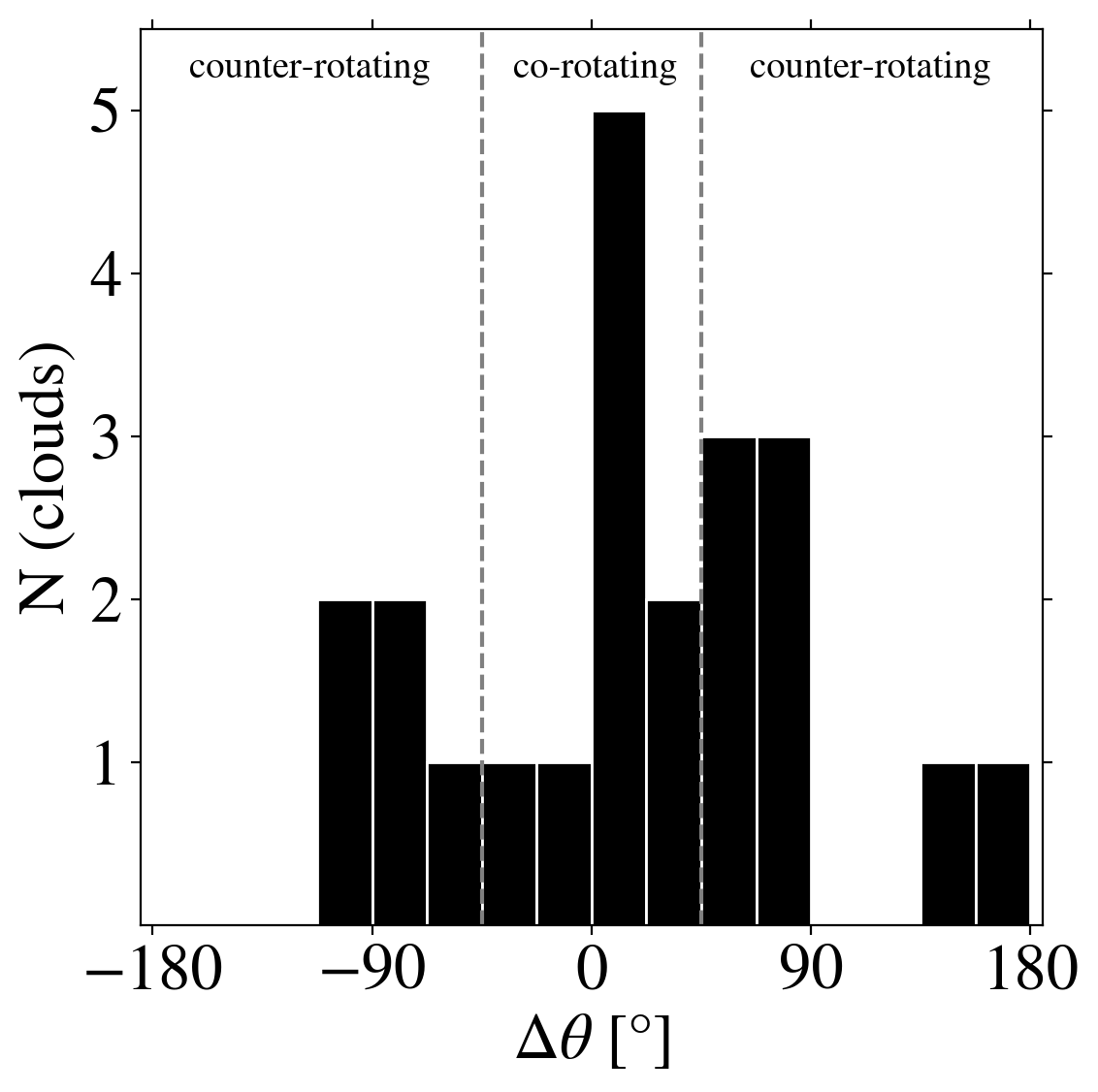}
    \caption{Distribution of angular separation between the rotation axis of molecular clouds and their corresponding HI envelopes.}
    \label{fig:H2vHI}
\end{figure}

In Figure \ref{fig:H2vHI}, we show the distribution of $\Delta \theta \equiv |\theta_\mathrm{H_2} - \theta_\mathrm{HI}|$, i.e., the separation between the rotation axes of molecular clouds and their HI envelopes. We find that 9 (40\%) of the molecular clouds are co-rotating, i.e.,  $\Delta \theta < 45\degree$, with respect to their HI envelope. Similarly, \cite{2011ApJ...732...79I} find that less than half of the molecular clouds are co-rotating with respect to their HI envelopes.

\newpage
\subsection{Goodness-of-fit of planar model}\label{sec:goodnessoffit}
\begin{figure}[ht!]
    \centering
    \includegraphics[width=\linewidth]{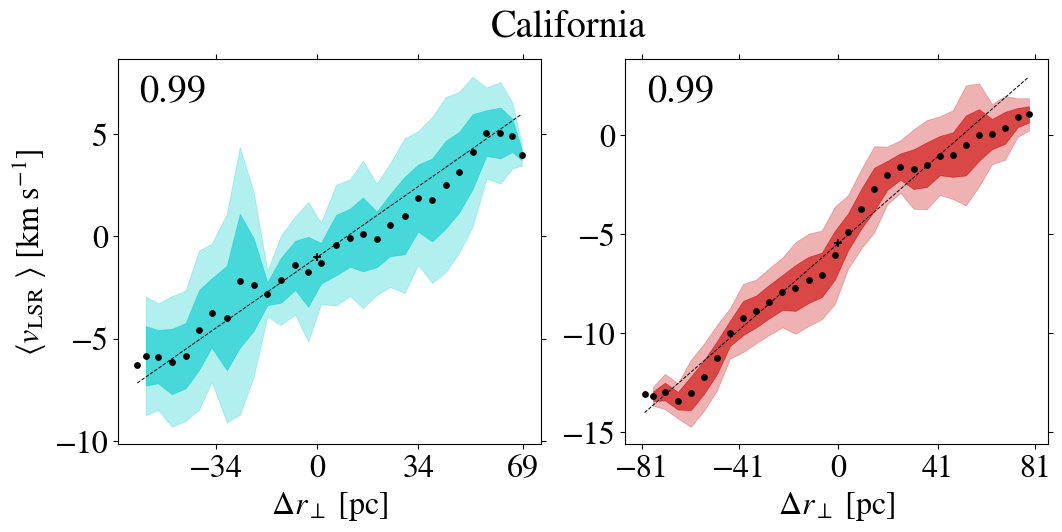}
    \caption{Intensity-weighted velocity centroid, $\langle v_\mathrm{LSR} \rangle$, as a function of the perpendicular displacement, $r_\perp$, from the rotation axis of the California molecular cloud (\textit{left}) and its HI envelope (\textit{right}). The dashed lines indicate the planar model, and the shaded regions show the $\pm 1\sigma$ and $\pm 2\sigma$ scatter of the velocity field map at each radial bin. The Pearson correlation coefficient is in the upper left corner of each plot. Velocity profiles of the remaining molecular clouds and HI envelopes can be found available in the Appendix \ref{sec:appndx_vvsr}.}.
    \label{fig:vvsr}
\end{figure}

\begin{figure}[h!]
    \centering
    \includegraphics[width=\linewidth]{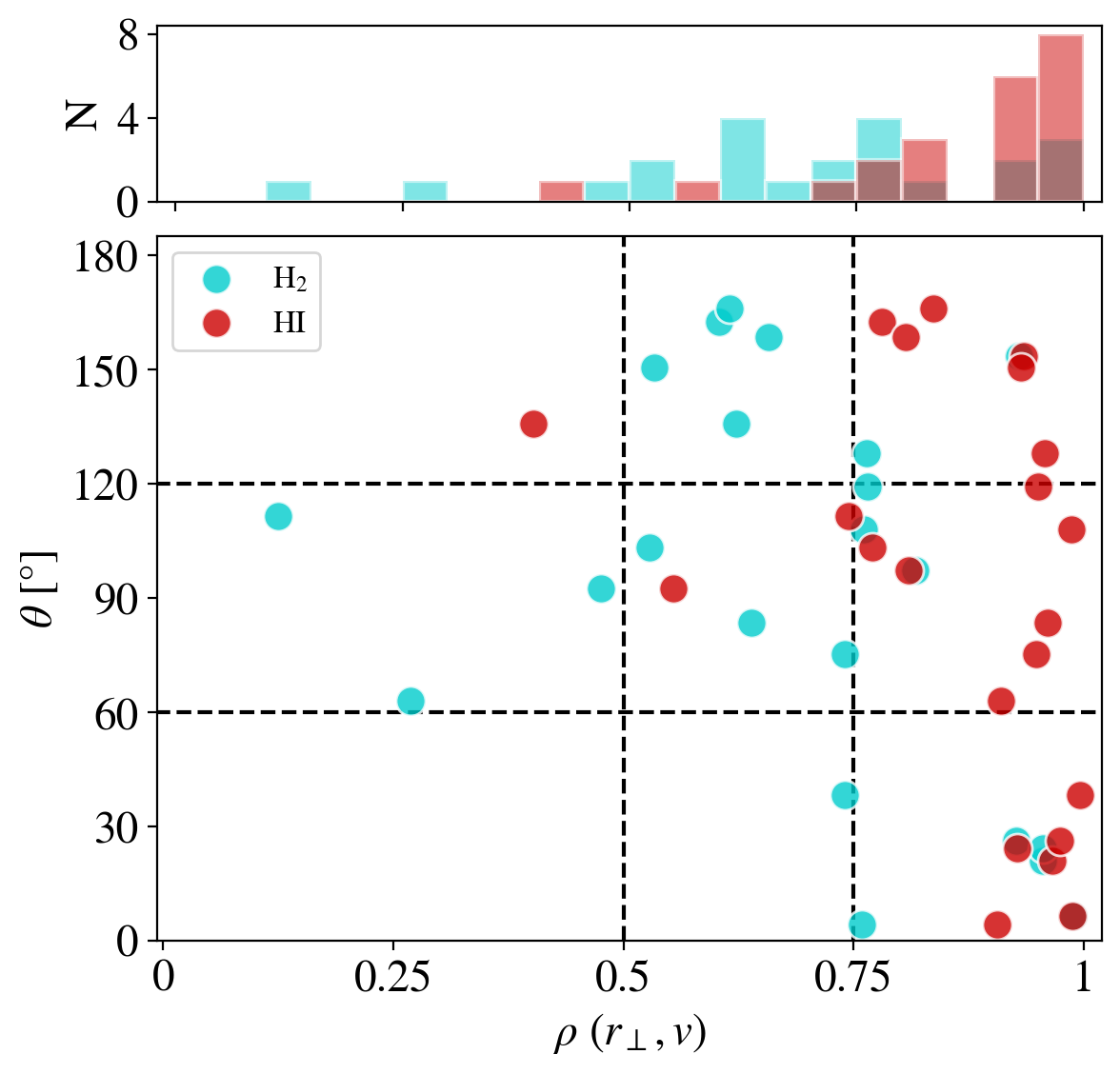}
    \caption{\textit{Top}: Distribution of Pearson correlation coefficients for the molecular clouds (cyan) and the HI envelopes (red). \textit{Below}: Velocity gradient directions as a function of Pearson coefficients. We observe highly linear gradients across all $\theta_\mathrm{HI}$, while molecular clouds show a moderate correlation between larger $\theta_\mathrm{H_2}$ and smaller Pearson coefficients.}
    \label{fig:pvals}
\end{figure}

In a planar velocity field, the line-of-sight velocity, $\langle v_\mathrm{LSR} \rangle$, will vary linearly with perpendicular distance from the rotation axis, $\Delta r_\perp$, with a slope equal to $\Omega$. 
To determine the extent to which the first moment maps of both populations clouds are planar, we plot the velocity profiles of both the molecular and atomic gas (Figure \ref{fig:vvsr}). For each velocity profile, we measure the Pearson correlation coefficient, $\rho(\Delta r_\perp, \langle v_\mathrm{LSR} \rangle)$. The distribution of $\rho$ for both populations is shown in Figure \ref{fig:pvals}. 

We find that 19 (86\%) of the HI envelopes have velocity profiles with correlation coefficients $\rho_\mathrm{HI}\geq 0.75$, of which 14 have $\rho_\mathrm{HI}>0.9$. Similarly, all but three of the molecular clouds have correlation coefficients greater than $\rho_\mathrm{H_2}\geq 0.5$, and 10 have above $0.75$. The molecular clouds have a median of $\rho_\mathrm{H_2}\approx 0.74$, while the HI envelopes have a nearly linear median correlation coefficient of $0.94$. This demonstrates that the velocity fields of both populations are reasonably well-approximated by a planar model, justifying the use of the large-scale 2D gradients to describe their kinematics. 

\section{Angular Momentum}\label{sec:j_trends}

The angular momentum of a rotating body is given by:

\begin{align}\label{eq:J}
    \overrightarrow{J} &= \overrightarrow{I}\ \times \overrightarrow{\omega} \approx \mathrm{c} M R^2\overrightarrow{\Omega},
\end{align}

where the angular velocity, $\omega$, can be replaced by the velocity gradient, $\Omega$, for solid-body rotation ($dv/dR = 0$), and $c$ is an order-unity coefficient set by the body's shape, rotation axis, and density distribution. We adopt a value of $c = 0.4$, applicable to a uniform density sphere. For an object whose primary rotation axis is in the plane of the sky, $\overrightarrow{\Omega}$ would appear as a linear gradient in the velocity field. 

The rotational angular momentum per unit mass is then given by:
\begin{align}
    \overrightarrow{j} \equiv \frac{\overrightarrow{J}}{M} = cR^2 \overrightarrow{\Omega}.
\end{align}

Note that the large-scale gradient in the cloud velocity fields need not be due to rotation; \cite{2000ApJ...543..822B} have shown that a statistically robust distribution of the angular momenta of of turbulent molecular clouds can be reproduced using the velocity gradient, as in the above equation. Below, we present our measurements of the specific angular momenta of the two phases, $j_\mathrm{H_2}$ and $j_\mathrm{HI}$. We then discuss how $j_\mathrm{H_2}$ and $j_\mathrm{HI}$ compare with each other, as well as where they land among observations in the literature.

\subsection{Specific angular momentum}

\begin{figure}
    \centering
    \includegraphics[width=\linewidth]{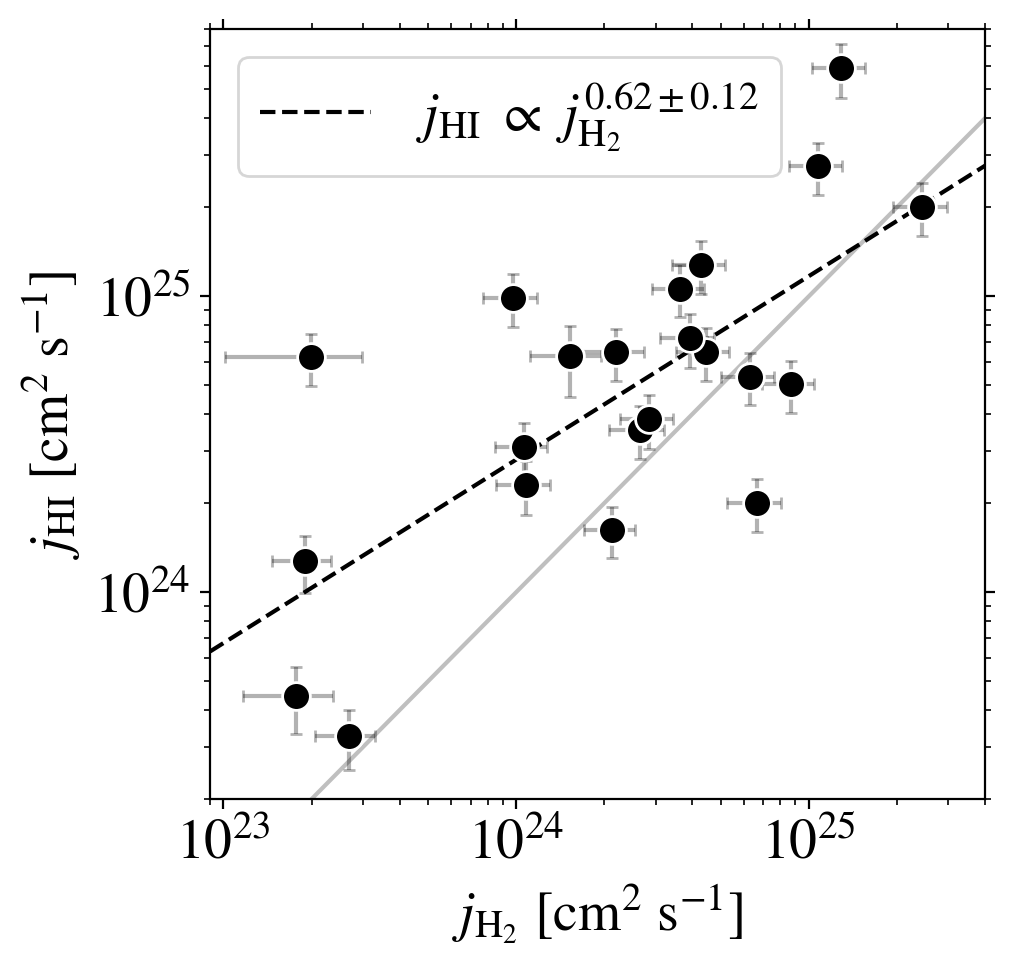}
    \caption{The specific angular momentum of HI envelopes with respect to those of their corresponding molecular clouds. The error bars are indicate the 1-$\sigma$ uncertainty of each measurement, derived from the distance uncertainties reported in \citep{2019ApJ...879..125Z} and the errors of fitting a plane to the first moment maps, reported in Table 2. The grey solid indicates where $j_\mathrm{HI} = j_\mathrm{H_2}$, and the black dashed line is the least-squares best fit to the data.}
    \label{fig:jj}
\end{figure}

The molecular clouds have specific angular momenta in the range $j_\mathrm{H_2} \in (1.8\times 10^{23},\ 2.4\times 10^{25})\ \mathrm{cm^2\ s^{-1}}$, with a median value of $2.7\times 10^{24}\ \mathrm{cm^2\ s^{-1}}$ (Table 2). On the other hand, the HI envelopes have $j_\mathrm{HI} \in (3.3\times 10^{23},\ 5.9\times 10^{25})\ \mathrm{cm^2\ s^{-1}}$, with a median of $5.8\times 10^{24}\ \mathrm{cm^2\ s^{-1}}$. We find that 17 of the systems have $j_\mathrm{HI}/j_\mathrm{H_2} >1$, and the ratio of the specific angular momenta of the two phases spans $0.3 - 31$, with an average of 4. This is similar to what IB measure for clouds in the Milky Way ($j_\mathrm{HI}/j_\mathrm{H_2} \approx 3$) and about a factor of 7 smaller than what IBB find for M33 clouds ($j_\mathrm{HI}/j_\mathrm{H_2} \approx 27$). The HI specific angular momenta scale as $j_\mathrm{HI} \propto j_\mathrm{H_2}^{0.62 \pm 0.12}$ (Figure \ref{fig:jj}), which is much steeper than the scaling IBB determine in M33, $j_\mathrm{HI} \propto j_\mathrm{H_2}^{0.17 \pm 0.05}$.

The angular momentum problem which IB demonstrated persists in the population level trends we find in this work. IB used masses calculated from the optically thick $^{12}\mathrm{CO}$ line. For the velocity fields, they used $^{13}\mathrm{CO}$, since the narrower linewidth enables finer velocity resolution. For each of the five clouds in common—Perseus, Orion A, Mon OB1, Mon R2, and Rosette—we measure masses that are from $\sim 1.3$ to 3.0 times higher than IB. 

IB measure the large-scale velocity gradients of the molecular clouds using $^{13}\mathrm{CO}$, while we use $^{12}\mathrm{CO}$ in this work. Our measurements for $\Omega_\mathrm{H_2}$ are consistent with one another for Monoceros R2, Perseus, and Rosette; on the other hand, we find that a nearly factor of 2 discrepancy for Monoceros OB1 and Orion A. The average  $j_\mathrm{HI}/j_\mathrm{H_2}$ we measure for the 5 clouds ($\sim 1.5$) is smaller than the average ratio that we get for the whole population as well as what IB calculate. This is because the Monoceros OB1 and Rosette systems are part of the 4 pairs where $j_\mathrm{HI} < j_\mathrm{H_2}$.

The above results, combined with the mis-alignment between the angular momentum vectors of molecular clouds and their HI envelopes (Figure \ref{fig:H2vHI}), argues that molecular cloud formation is more complicated than top-down, angular-momentum-conserving collapse of atomic gas. Instead, the observations suggest that angular momentum is redistributed as the molecular cloud forms. Alternatively, collapse may occur along multiple spatial dimensions simultaneously, and our model does not account for the third dimension into the plane of the sky---or mostly likely, a combination of both.
\begin{figure}[ht!]
    \centering
    \includegraphics[width=\linewidth]{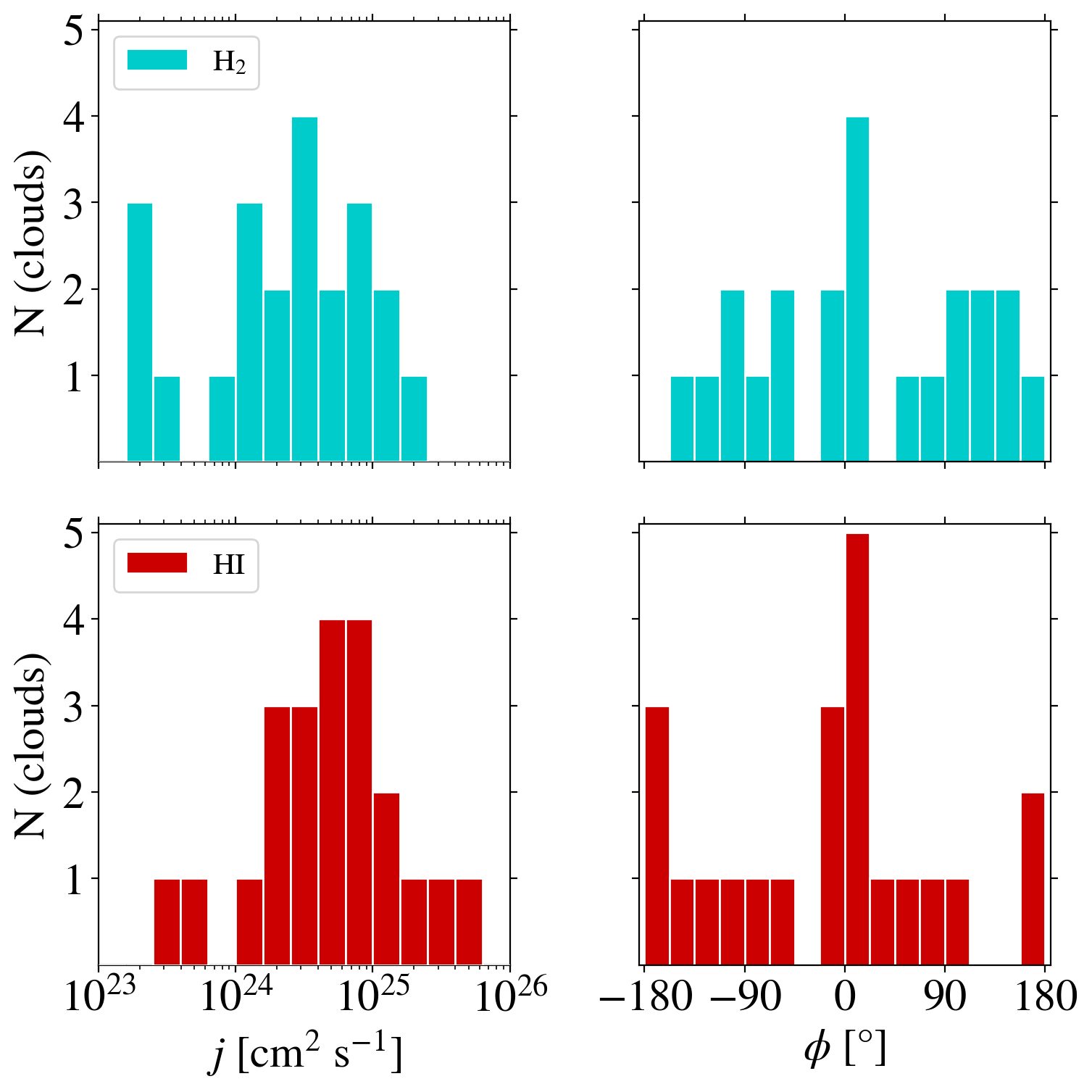}
    \caption{Kinematic properties of the molecular clouds (first row) and HI envelopes (second row). The magnitude and direction of the specific angular momenta of each population are shown in the left and right columns, respectively.}
    \label{fig:init_stat}
\end{figure}

\subsection{Scaling Relations}
Turbulence plays a key role in supporting molecular clouds against gravitational collapse (e.g., \citealt{1981MNRAS.194..809L}). This turbulence is primarily supersonic, following a size-linewidth scaling of $\sigma \propto R^{1/2}$ \citep{2000ApJ...543..822B}. The angular velocity of a cloud and its linewidth scale as $\Omega \propto \sigma/R$, implying a specific angular momentum-size relation of $j \propto R^{3/2}$ \citep{2000ApJ...543..822B}.  

\begin{figure*}
    \centering
    \includegraphics[width=\linewidth]{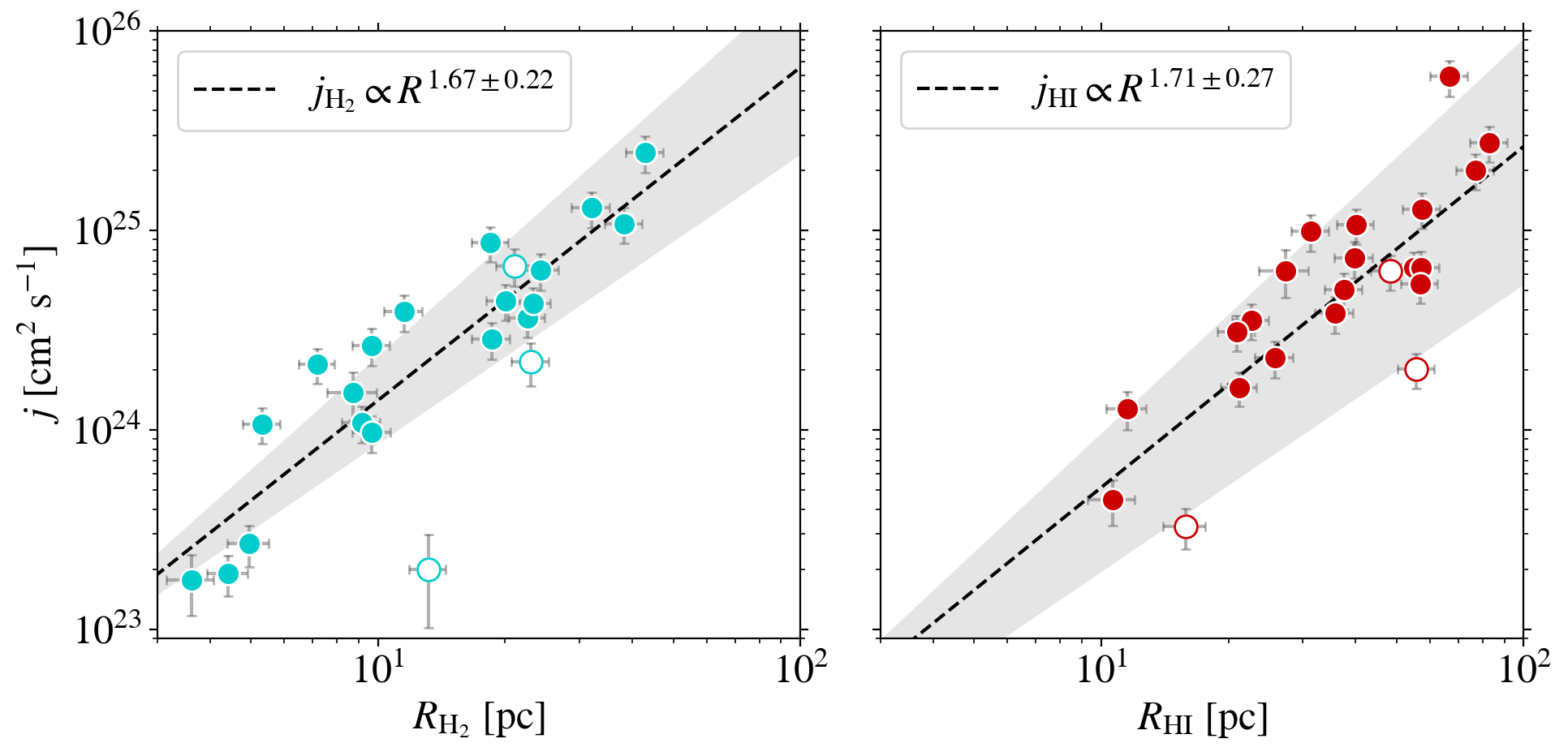}
    \caption{The specific angular momentum of molecular clouds (\textit{left}) and their HI envelopes (\textit{right}) as a function of size. The dashed lines are the least-squares fit to the data, and the shaded regions indicate the $\pm 1\sigma$ error of each fit. The unfilled data points belong to molecular clouds (HI envelopes) with correlation coefficients of $\leq 0.5$ (0.75).}
    \label{fig:jRs}
    
\end{figure*}

Least-squares fitting to our measurements shows that the molecular clouds have specific angular momenta that scale as $j_\mathrm{H_2} \propto R^{1.67\pm0.22}$. The HI envelope follow a steeper scaling $j_\mathrm{HI} \propto R^{1.71\pm0.27}$. Both are within $1\sigma$ of the $j \propto R^{1.5}$ relation expected of a supersonic turbulent cascade (Figure \ref{fig:jRs}).

\begin{figure}
    \centering
    \includegraphics[width=\linewidth]{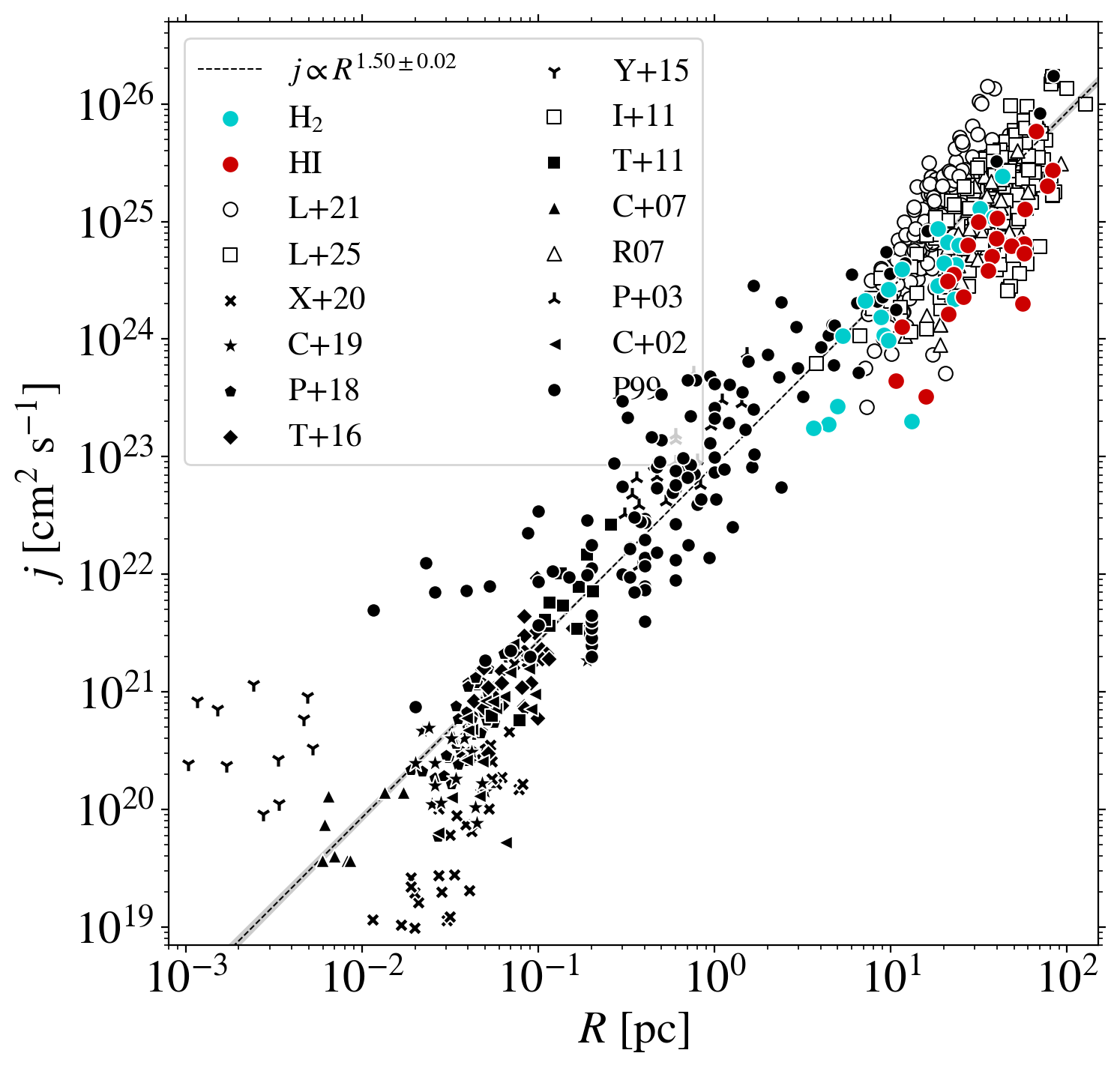}
    \caption{Size-Angular Momentum relation of star-forming regions in the Solar Neighborhood. Extragalactic observations are shown with unfilled markers. The magenta line shows the least-squares fit to the combined data; a supersonic turbulent cascade is observed in the size range $10^{-3} <  R < 10^2\ \mathrm{pc}$. }
    \label{fig:jRs_all}  
    
\end{figure}

To examine the evolution of specific angular momentum across size scales---from the substructure in molecular clouds, to the atomic envelopes encompassing them---we combine our results with previous observations of the kinematics of star-forming regions. We draw from a database compiled by \cite{1999A&AS..134..241P} of observations of molecular clouds and cores spanning across three decades, utilizing a diversity of tracers \citep{1998yCat..41340241P}. We also include recent observations conducted by  \cite{2021MNRAS.505.4048L,Xu__2020,2019ApJ...886..119C,2018A&A...617A..27P,2016PASJ...68...24T,2015ApJ...799..193Y,2011ApJ...732...79I,2011ApJ...740...45T,2007ApJ...669.1058C,2003A&A...405..639P,2002ApJ...572..238C}. Similar meta-analyses have been done in the past  for observational (e.g., \citealt{2023MNRAS.525..364P,2019MNRAS.490..527C,1999A&AS..134..241P}) and simulated cloud populations (e.g., \citealt{2026arXiv260111830A,2018ApJ...865...34C}). In this work we combine for the first time observational data over 5 orders of magnitude in size, including extragalactic clouds. To facilitate comparison, we calculate all specific angular momenta using the geometric coefficient of 0.4, even when the original work does otherwise. Where there is a major and minor axis reported for the spatial extent of an object, we calculate the radius as the geometric mean of the two. 

With the results from this study, we show that molecular clouds and cores in the size range $R\in (10^{-3},\ 10^2)$ pc follow a power-law relationship of $j\propto R^{1.50 \pm 0.02}$ (Figure \ref{fig:jRs_all}). This suggests that angular momentum evolves uniformly with size throughout the ISM from dense pre-stellar cores to diffuse atomic clouds.


\section{Discussion}\label{sec:discussion}

We now describe a simple physical model describing angular momentum transport that occurs during the formation of a molecular cloud. We calculate the properties of the progenitor atomic gas from which a molecular cloud forms. We also calculate the timescale for angular momentum redistribution and analyze the motion of HI envelopes within the context of the Galaxy.

\newpage
\subsection{Estimating momentum transport timescales}\label{sec:tdis} Magnetohydrodynamic simulations (e.g., \citealt{2024ApJ...963..106M,2022ApJ...925...78A}) demonstrate that molecular clouds lose angular momentum over time as they form. \cite{2022ApJ...925...78A} find that as the gas undergoes hierarchical gravitational collapse, angular momentum  is transported from the dense, contracting molecular cloud into the surrounding diffuse atomic gas. Below, we describe the molecular cloud and HI envelope as a co-evolving system with a shared common progenitor, in order to estimate the timescale for angular momentum transport in molecular clouds.

Let us imagine a progenitor HI cloud with initial angular momentum $J_\mathrm{init}$ that, due to gravitational collapse or shocks, increases in density sufficiently to form a molecular cloud in its interior. 
The present-day HI we observe will be the gas that remained atomic and has co-evolved with the molecular cloud. The total angular momentum of this system would be:\textbf{
}
\begin{align}
   \overrightarrow{J_\mathrm{total}} = \overrightarrow{J_\mathrm{init}} + \overrightarrow{\dot{J}}  = (\overrightarrow{J_\mathrm{HI}} + \overrightarrow{J_\mathrm{H_2}})_\mathrm{present}
\end{align} 
 
where $J_\mathrm{HI}$ and $J_\mathrm{H_2}$ are the total angular momenta of the present HI envelope and molecular cloud respectively, $J_\mathrm{init}$ is the total angular momentum of the progenitor, and $\dot{J}$ is the change in angular momentum due to interactions with the environment, e.g., inflowing or outflowing gas. In the simplest scenario of an isolated system, $\dot{J} = 0$ and $M_\mathrm{tot} = M_\mathrm{init} = M_\mathrm{HI} + M_\mathrm{H_2}$. The specific angular momentum of the progenitor HI cloud can then be written as: 

\begin{align}
M_\mathrm{H_2}\overrightarrow{j_\mathrm{H_2}} + M_\mathrm{HI}\overrightarrow{j_\mathrm{HI}} &=  M_\mathrm{tot}\overrightarrow{j_\mathrm{init}}
\end{align}  

Writing out the components of the vectors explicitly, we get:
\begin{align}
    \frac{c}{M_\mathrm{tot}} \sum M_hR_h^2\binom{\Omega_x}{\Omega_y}_h &=  \binom{j_x}{j_y}_\mathrm{init}.
\end{align}

where the index $h$ in the sum goes over the molecular cloud and HI envelope, and we have assumed that both cloud phases are described by the same geometric factor $c$. While the total angular momentum is a 3-dimensional vector, we restrict our calculations to the two on-sky components, $x$ and $y$, accessible from observations.

The progenitor HI clouds of the systems we study here have angular momenta of $j_\mathrm{init} \in (2.8 \times 10^{23},\ 4.5\times 10^{25})\ \mathrm{cm^2\ s^{-1}}$, with a median of $4.2\times 10^{24}\ \mathrm{cm^2\ s^{-1}}$. The progenitor clouds have specific angular momenta that are $\approx 3$ times larger than $j_\mathrm{H_2}$ and $0.8\times j_\mathrm{HI}$, which puts them as intermediate between the present day molecular cloud and HI envelope. This suggests that the increase from $j_\mathrm{init}$ to $j_\mathrm{HI}$ is caused by the molecular cloud redistributing angular momentum into the surrounding HI, consistent with IBB's measurements of isolated HI clouds having smaller angular momenta than those associated with molecular clouds. 

To estimate the timescale over which the angular momentum of the progenitor system is transported away from the molecular cloud, we assume that the dominant torque for redistributing angular momentum is also responsibly for dissipating the rotational energy from the system. By dimensional analysis, the angular momentum redistribution rate can be written as $ {\Delta J} / {\Delta t_\mathrm{diss}} \approx \Delta E$. We then express the redistribution timescale $\Delta t_\mathrm{diss}$ in terms of observationally-derived quantities:
\begin{align}\label{eq:delta_t}
    \Delta t_\mathrm{diss} &= \frac{J_\mathrm{HI}}{E_\mathrm{HI}} = \frac{j_\mathrm{HI}}{e_\mathrm{HI}}     
\end{align}  
Above, we have assumed that total change in angular momentum $\Delta J$ reduces to $\Delta J = J_\mathrm{init} - J_\mathrm{H_2} = J_\mathrm{HI}$, and that the same logic applies to the total change in rotational energy: $\Delta E = E_\mathrm{HI}$. The specific rotational energy of the HI envelope is given by:
\begin{align}
     e_\mathrm{HI} \approx  (\Omega R)^2 
\end{align}

The redistribution timescales we calculate using Equation \ref{eq:delta_t} are $\Delta t_\mathrm{dis} \in (4,\ 80)\ \mathrm{Myr}$\footnote{All but Monoceros OB1 have $\Delta t_\mathrm{dis} < 40\ \mathrm{Myr}$, which has a longer timescale of $\sim 80\ \mathrm{Myr}$}, with an overall median of $13\ \mathrm{Myr}$. This is similar to the free-fall time of the HI envelopes ($\sim 14\ \mathrm{Myr})$, and twice as long as the average sound crossing time of both populations and the free-fall time of the molecular clouds ($\sim 6\ \mathrm{Myr}$). The redistribution timescale is also 5 (15) times shorter than typical molecular cloud (HI envelope) rotational period. This suggests that potential physical mechanisms for redistributing the angular momentum---e.g., magnetic braking, turbulent dissipation, gravitational torques---will have had the opportunity to brake the gas as it contracts, facilitating gravitational collapse.

\begin{figure}[h!]
\includegraphics[width=\linewidth]{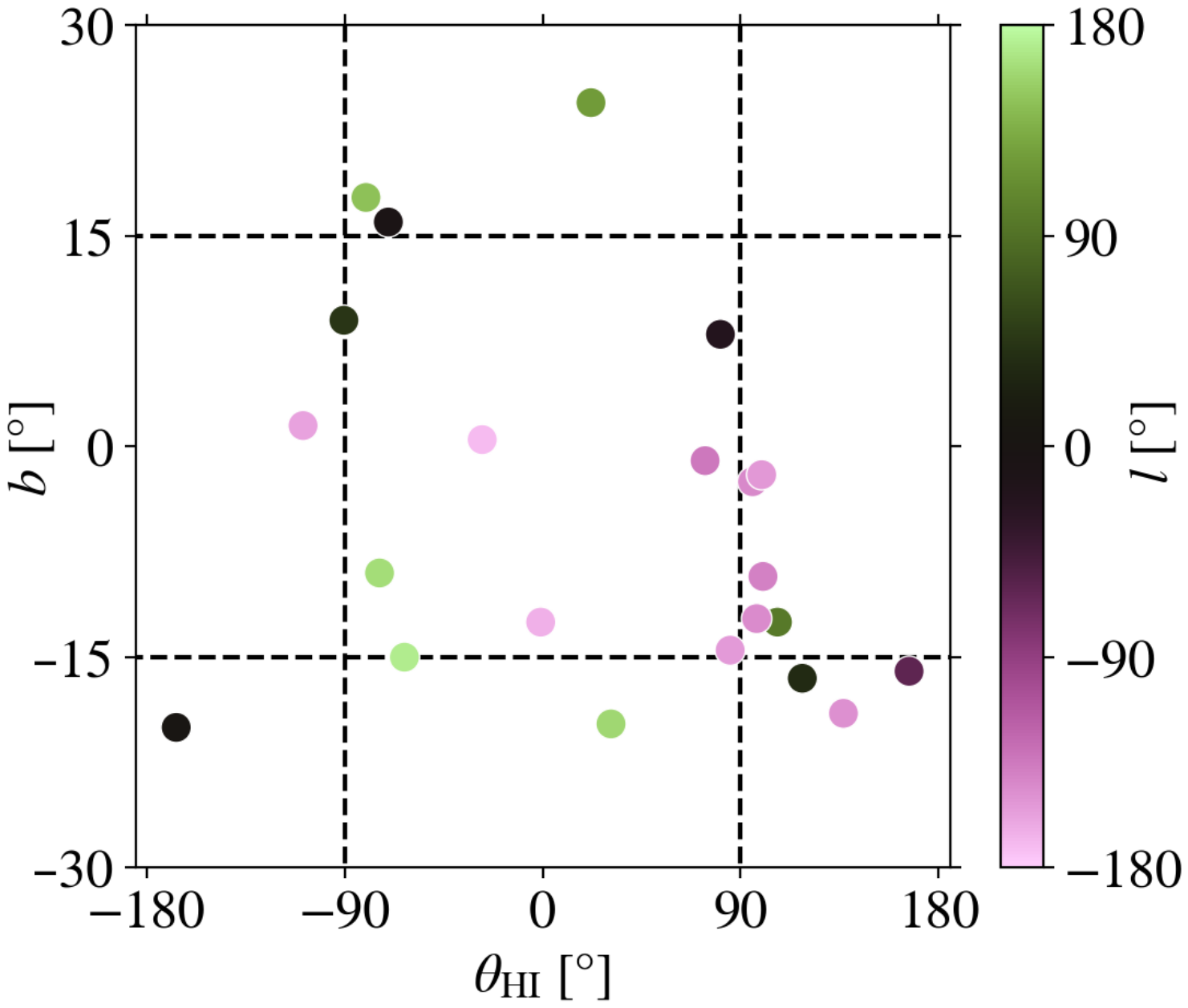}
\caption{Velocity gradient position angles of the HI envelopes, arranged according to Galactic latitude. The colorbar indicates the Galactic longitude coordinate of each envelope.}
\label{fig:thetaHI}
\end{figure}

\subsection{Overlapping HI envelopes}
The axis of large-scale Galaxy rotation lies along $\phi = 90\degree$ when viewed from the midplane. About $27\%$ (6) of the molecular clouds, and 3 of the HI envelopes have angular momentum vectors aligned with the Milky Way $\Delta \phi = |\phi_\mathrm{Gal} - \phi_\mathrm{H_2}| < 45\degree$ (Figure \ref{fig:init_stat}). On the other hand, three times as many of the HI clouds have $|\phi| \leq 45 \degree$. If $\Omega_\mathrm{HI}$ arose due to rotation, this would imply that their rotation planes that are perpendicular to the Galactic plane, which is unexpected given the dynamical coupling of HI clouds to the Galaxy (e.g., \citealt{Jeffreson_2020}), this is unexpected. One would rather expect that $\phi_\mathrm{HI} \sim 90\degree$. Further examination shows that 5 of these HI envelopes share similar on-sky coordinates (Figure \ref{fig:thetaHI}), with the overlap among their velocity fields artificially increasing the occurrence rate of vertical velocity gradients. 

Future work can improve upon this by distinguishing between molecular clouds that share a common HI envelope, and those where the overlap is due to chance. The former will require refining our model of angular momentum redistribution to account for the co-evolution of multiple molecular clouds within a shared envelope. For those clouds whose HI envelopes overlap purely due to projection, the velocity fields of the foreground and background systems need to be disentangled. This necessitates characterizing the distribution of emitting material along the line-of-sight, e.g., using dust-derived three-dimensional reconstructions of the Solar Neighborhood such as those by \citealt{2024MNRAS.532.3480D,2024A&A...685A..82E,2020A&A...639A.138L}.

\section{Summary and conclusions}\label{sec:summary}
With the aim of investigating the angular momentum in molecular clouds and their associated HI envelopes, we identify atomic gas associated with 22 molecular clouds in the Solar Neighborhood, following \cite{2011ApJ...732...78I} and \cite{2011ApJ...732...79I}. 
\begin{itemize}
\item We measure the properties of the molecular clouds and their associated HI envelopes using $^{12}\mathrm{CO}$ and 21-cm observations. The accumulation radii of the HI envelopes are on average 3 times larger and more massive compared to the molecular cloud with which they are associated (Table 2).
\item The velocity field maps (Figure \ref{fig:maps}) of both populations are fit with planes to derive the large-scale velocity gradients, $\Omega_\mathrm{HI}$ and $\Omega_\mathrm{H_2}$. The molecular clouds have typical velocity gradients of $\Omega_\mathrm{H_2} \approx 0.1 \ \mathrm{km\ s^{-1}\ pc^{-1}}$, while the HI have gradients of $\Omega_\mathrm{HI} \approx 0.03\ \mathrm{km\ s^{-1}}$. The rotation axes of a molecular clouds and its HI envelopes tend to be unaligned, with separations distributed randomly between $0\degree$ and $180\degree$ (Figure \ref{fig:H2vHI}).
\item Our results demonstrate that a simple top-down collapse model is insufficient to explain the discrepancy between the angular momenta of molecular clouds and those of the atomic gas in their vicinity. The average specific angular momentum of the HI envelopes is $j_\mathrm{HI} \approx 9.1 \times 10^{24}\ \mathrm{cm^2\ s^{-1}}$, and is on average 4 times larger than the corresponding $j_\mathrm{H_2}$ (Figure \ref{fig:jj}). Along with the abundance of systems with large angular separations of molecular cloud and HI rotation axes, our results builds on a growing body of evidence that molecular clouds have systematically smaller, and misaligned, angular momenta than would be expected had they formed via direct collapse from the surrounding HI. This result suggests that angular momentum redistribution has occurred following the formation of molecular clouds. 
\item The specific angular momentum of molecular clouds scales with size as $j_\mathrm{H_2} \propto R^{1.67 \pm 0.22}$, while the HI envelopes follow a steeper relation of $j_\mathrm{HI} \propto R^{1.71 \pm 0.27}$ (Figure \ref{fig:jRs}). Both are consistent with the $j \propto R^{1.5}$ scaling expected of supersonic turbulence. We also demonstrate that molecular clouds and cores over five orders of magnitude in size follow the $j\propto R^{1.5}$ scaling of a supersonic turbulent cascade (Figure \ref{fig:jRs_all}).
\item We develop a physical model describing the coevolution of the angular momenta of the molecular and atomic phases in Section \ref{sec:tdis}, and estimate the specific angular momentum of the progenitor HI cloud from which the present-day molecular cloud-HI envelope system formed. We use also this model to derive the timescales on which angular momentum is redistributed from the molecular clouds to their surrounding HI. We find that the redistribution timescales, $\Delta t_\mathrm{dis} \sim 13\ \mathrm{Myr}$, are within a factor of 2 of the sound crossing time and free-fall time of the clouds.

\end{itemize}

\begin{acknowledgements}
We are grateful to Eric Koch, John Forbes, Chris McKee, and Karin Sandstrom for insightful discussions that helped to improve this paper. We also thank an anonymous referee for their constructive feedback. We acknowledge support from the Heising–Simons Foundation (grant 2022-3532) for this work.
\end{acknowledgements}

\software{astropy \citep{astropy:2013,astropy:2018,astropy:2022},  
         matplotlib \citep{2016zndo.....61948D},
         pandas \citep{2022zndo...6053272R},
         scipy \citep{2018zndo...1218715V},
         numpy \citep{harris2020array},
         joblib \citep{joblib_development_team_2025_15496554}}
      
\bibliography{references}

@article{2016zndo.....61948D,
       author = {{Developers}, Matplotlib},
        title = "{matplotlib: v1.5.3}",
         year = 2016,
        month = sep,
          eid = {10.5281/zenodo.61948},
          doi = {10.5281/zenodo.61948},
      version = {v1.5.3},
    publisher = {Zenodo},
       adsurl = {https://ui.adsabs.harvard.edu/abs/2016zndo.....61948D}
}

@ARTICLE{2008ApJ...677..327M,
       author = {{Machida}, Masahiro N. and {Tomisaka}, Kohji and {Matsumoto}, Tomoaki and {Inutsuka}, Shu-ichiro},
        title = "{Formation Scenario for Wide and Close Binary Systems}",
      journal = {\apj},
         year = 2008,
        month = apr,
       volume = {677},
       number = {1},
        pages = {327-347},
          doi = {10.1086/529133},
archivePrefix = {arXiv},
       eprint = {0709.2739},
 primaryClass = {astro-ph},
       adsurl = {https://ui.adsabs.harvard.edu/abs/2008ApJ...677..327M}
}

@ARTICLE{1984MNRAS.206..197L,
author = {{Larson}, R.~B.},
title = "{Gravitational torques and star formation}",
journal = {\mnras},
year = 1984,
month = jan,
volume = {206},
pages = {197-207},
doi = {10.1093/mnras/206.1.197},
adsurl = {https://ui.adsabs.harvard.edu/abs/1984MNRAS.206..197L}
}

@ARTICLE{2024ApJ...963..106M,
author = {{Misugi}, Yoshiaki and {Inutsuka}, Shu-ichiro and {Arzoumanian}, Doris and {Tsukamoto}, Yusuke},
title = "{Evolution of the Angular Momentum of Molecular Cloud Cores in Magnetized Molecular Filaments}",
journal = {\apj},
year = 2024,
month = mar,
volume = {963},
number = {2},
eid = {106},
pages = {106},
doi = {10.3847/1538-4357/ad1990},
archivePrefix = {arXiv},
eprint = {2312.16920},
primaryClass = {[astro-ph.GA](http://astro-ph.ga/)},
adsurl = {https://ui.adsabs.harvard.edu/abs/2024ApJ...963..106M}
}

@article{astropy:2013,
Adsurl = {http://adsabs.harvard.edu/abs/2013A%26A...558A..33A},
Archiveprefix = {arXiv},
Author = {{Astropy Collaboration} and {Robitaille}, T.~P. and {Tollerud}, E.~J. and {Greenfield}, P. and {Droettboom}, M. and {Bray}, E. and {Aldcroft}, T. and {Davis}, M. and {Ginsburg}, A. and {Price-Whelan}, A.~M. and {Kerzendorf}, W.~E. and {Conley}, A. and {Crighton}, N. and {Barbary}, K. and {Muna}, D. and {Ferguson}, H. and {Grollier}, F. and {Parikh}, M.~M. and {Nair}, P.~H. and {Unther}, H.~M. and {Deil}, C. and {Woillez}, J. and {Conseil}, S. and {Kramer}, R. and {Turner}, J.~E.~H. and {Singer}, L. and {Fox}, R. and {Weaver}, B.~A. and {Zabalza}, V. and {Edwards}, Z.~I. and {Azalee Bostroem}, K. and {Burke}, D.~J. and {Casey}, A.~R. and {Crawford}, S.~M. and {Dencheva}, N. and {Ely}, J. and {Jenness}, T. and {Labrie}, K. and {Lim}, P.~L. and {Pierfederici}, F. and {Pontzen}, A. and {Ptak}, A. and {Refsdal}, B. and {Servillat}, M. and {Streicher}, O.},
Doi = {10.1051/0004-6361/201322068},
Eid = {A33},
Eprint = {1307.6212},
Journal = {\aap},
Month = oct,
Pages = {A33},
Primaryclass = {astro-ph.IM},
Title = {{Astropy: A community Python package for astronomy}},
Volume = 558,
Year = 2013}

@ARTICLE{astropy:2018,
       author = {{Astropy Collaboration} and {Price-Whelan}, A.~M. and
         {Sip{\H{o}}cz}, B.~M. and {G{\"u}nther}, H.~M. and {Lim}, P.~L. and
         {Crawford}, S.~M. and {Conseil}, S. and {Shupe}, D.~L. and
         {Craig}, M.~W. and {Dencheva}, N. and {Ginsburg}, A. and {Vand
        erPlas}, J.~T. and {Bradley}, L.~D. and {P{\'e}rez-Su{\'a}rez}, D. and
         {de Val-Borro}, M. and {Aldcroft}, T.~L. and {Cruz}, K.~L. and
         {Robitaille}, T.~P. and {Tollerud}, E.~J. and {Ardelean}, C. and
         {Babej}, T. and {Bach}, Y.~P. and {Bachetti}, M. and {Bakanov}, A.~V. and
         {Bamford}, S.~P. and {Barentsen}, G. and {Barmby}, P. and
         {Baumbach}, A. and {Berry}, K.~L. and {Biscani}, F. and {Boquien}, M. and
         {Bostroem}, K.~A. and {Bouma}, L.~G. and {Brammer}, G.~B. and
         {Bray}, E.~M. and {Breytenbach}, H. and {Buddelmeijer}, H. and
         {Burke}, D.~J. and {Calderone}, G. and {Cano Rodr{\'\i}guez}, J.~L. and
         {Cara}, M. and {Cardoso}, J.~V.~M. and {Cheedella}, S. and {Copin}, Y. and
         {Corrales}, L. and {Crichton}, D. and {D'Avella}, D. and {Deil}, C. and
         {Depagne}, {\'E}. and {Dietrich}, J.~P. and {Donath}, A. and
         {Droettboom}, M. and {Earl}, N. and {Erben}, T. and {Fabbro}, S. and
         {Ferreira}, L.~A. and {Finethy}, T. and {Fox}, R.~T. and
         {Garrison}, L.~H. and {Gibbons}, S.~L.~J. and {Goldstein}, D.~A. and
         {Gommers}, R. and {Greco}, J.~P. and {Greenfield}, P. and
         {Groener}, A.~M. and {Grollier}, F. and {Hagen}, A. and {Hirst}, P. and
         {Homeier}, D. and {Horton}, A.~J. and {Hosseinzadeh}, G. and {Hu}, L. and
         {Hunkeler}, J.~S. and {Ivezi{\'c}}, {\v{Z}}. and {Jain}, A. and
         {Jenness}, T. and {Kanarek}, G. and {Kendrew}, S. and {Kern}, N.~S. and
         {Kerzendorf}, W.~E. and {Khvalko}, A. and {King}, J. and {Kirkby}, D. and
         {Kulkarni}, A.~M. and {Kumar}, A. and {Lee}, A. and {Lenz}, D. and
         {Littlefair}, S.~P. and {Ma}, Z. and {Macleod}, D.~M. and
         {Mastropietro}, M. and {McCully}, C. and {Montagnac}, S. and
         {Morris}, B.~M. and {Mueller}, M. and {Mumford}, S.~J. and {Muna}, D. and
         {Murphy}, N.~A. and {Nelson}, S. and {Nguyen}, G.~H. and
         {Ninan}, J.~P. and {N{\"o}the}, M. and {Ogaz}, S. and {Oh}, S. and
         {Parejko}, J.~K. and {Parley}, N. and {Pascual}, S. and {Patil}, R. and
         {Patil}, A.~A. and {Plunkett}, A.~L. and {Prochaska}, J.~X. and
         {Rastogi}, T. and {Reddy Janga}, V. and {Sabater}, J. and
         {Sakurikar}, P. and {Seifert}, M. and {Sherbert}, L.~E. and
         {Sherwood-Taylor}, H. and {Shih}, A.~Y. and {Sick}, J. and
         {Silbiger}, M.~T. and {Singanamalla}, S. and {Singer}, L.~P. and
         {Sladen}, P.~H. and {Sooley}, K.~A. and {Sornarajah}, S. and
         {Streicher}, O. and {Teuben}, P. and {Thomas}, S.~W. and
         {Tremblay}, G.~R. and {Turner}, J.~E.~H. and {Terr{\'o}n}, V. and
         {van Kerkwijk}, M.~H. and {de la Vega}, A. and {Watkins}, L.~L. and
         {Weaver}, B.~A. and {Whitmore}, J.~B. and {Woillez}, J. and
         {Zabalza}, V. and {Astropy Contributors}},
        title = "{The Astropy Project: Building an Open-science Project and Status of the v2.0 Core Package}",
      journal = {\aj},
         year = 2018,
        month = sep,
       volume = {156},
       number = {3},
          eid = {123},
        pages = {123},
          doi = {10.3847/1538-3881/aabc4f},
archivePrefix = {arXiv},
       eprint = {1801.02634},
 primaryClass = {astro-ph.IM},
       adsurl = {https://ui.adsabs.harvard.edu/abs/2018AJ....156..123A}
}

@ARTICLE{astropy:2022,
       author = {{Astropy Collaboration} and {Price-Whelan}, Adrian M. and {Lim}, Pey Lian and {Earl}, Nicholas and {Starkman}, Nathaniel and {Bradley}, Larry and {Shupe}, David L. and {Patil}, Aarya A. and {Corrales}, Lia and {Brasseur}, C.~E. and {N{"o}the}, Maximilian and {Donath}, Axel and {Tollerud}, Erik and {Morris}, Brett M. and {Ginsburg}, Adam and {Vaher}, Eero and {Weaver}, Benjamin A. and {Tocknell}, James and {Jamieson}, William and {van Kerkwijk}, Marten H. and {Robitaille}, Thomas P. and {Merry}, Bruce and {Bachetti}, Matteo and {G{"u}nther}, H. Moritz and {Aldcroft}, Thomas L. and {Alvarado-Montes}, Jaime A. and {Archibald}, Anne M. and {B{'o}di}, Attila and {Bapat}, Shreyas and {Barentsen}, Geert and {Baz{'a}n}, Juanjo and {Biswas}, Manish and {Boquien}, M{'e}d{'e}ric and {Burke}, D.~J. and {Cara}, Daria and {Cara}, Mihai and {Conroy}, Kyle E. and {Conseil}, Simon and {Craig}, Matthew W. and {Cross}, Robert M. and {Cruz}, Kelle L. and {D'Eugenio}, Francesco and {Dencheva}, Nadia and {Devillepoix}, Hadrien A.~R. and {Dietrich}, J{"o}rg P. and {Eigenbrot}, Arthur Davis and {Erben}, Thomas and {Ferreira}, Leonardo and {Foreman-Mackey}, Daniel and {Fox}, Ryan and {Freij}, Nabil and {Garg}, Suyog and {Geda}, Robel and {Glattly}, Lauren and {Gondhalekar}, Yash and {Gordon}, Karl D. and {Grant}, David and {Greenfield}, Perry and {Groener}, Austen M. and {Guest}, Steve and {Gurovich}, Sebastian and {Handberg}, Rasmus and {Hart}, Akeem and {Hatfield-Dodds}, Zac and {Homeier}, Derek and {Hosseinzadeh}, Griffin and {Jenness}, Tim and {Jones}, Craig K. and {Joseph}, Prajwel and {Kalmbach}, J. Bryce and {Karamehmetoglu}, Emir and {Ka{l}uszy{'n}ski}, Miko{l}aj and {Kelley}, Michael S.~P. and {Kern}, Nicholas and {Kerzendorf}, Wolfgang E. and {Koch}, Eric W. and {Kulumani}, Shankar and {Lee}, Antony and {Ly}, Chun and {Ma}, Zhiyuan and {MacBride}, Conor and {Maljaars}, Jakob M. and {Muna}, Demitri and {Murphy}, N.~A. and {Norman}, Henrik and {O'Steen}, Richard and {Oman}, Kyle A. and {Pacifici}, Camilla and {Pascual}, Sergio and {Pascual-Granado}, J. and {Patil}, Rohit R. and {Perren}, Gabriel I. and {Pickering}, Timothy E. and {Rastogi}, Tanuj and {Roulston}, Benjamin R. and {Ryan}, Daniel F. and {Rykoff}, Eli S. and {Sabater}, Jose and {Sakurikar}, Parikshit and {Salgado}, Jes{'u}s and {Sanghi}, Aniket and {Saunders}, Nicholas and {Savchenko}, Volodymyr and {Schwardt}, Ludwig and {Seifert-Eckert}, Michael and {Shih}, Albert Y. and {Jain}, Anany Shrey and {Shukla}, Gyanendra and {Sick}, Jonathan and {Simpson}, Chris and {Singanamalla}, Sudheesh and {Singer}, Leo P. and {Singhal}, Jaladh and {Sinha}, Manodeep and {Sip{H{o}}cz}, Brigitta M. and {Spitler}, Lee R. and {Stansby}, David and {Streicher}, Ole and {{{S}}umak}, Jani and {Swinbank}, John D. and {Taranu}, Dan S. and {Tewary}, Nikita and {Tremblay}, Grant R. and {Val-Borro}, Miguel de and {Van Kooten}, Samuel J. and {Vasovi{'c}}, Zlatan and {Verma}, Shresth and {de Miranda Cardoso}, Jos{'e} Vin{'i}cius and {Williams}, Peter K.~G. and {Wilson}, Tom J. and {Winkel}, Benjamin and {Wood-Vasey}, W.~M. and {Xue}, Rui and {Yoachim}, Peter and {Zhang}, Chen and {Zonca}, Andrea and {Astropy Project Contributors}},
        title = "{The Astropy Project: Sustaining and Growing a Community-oriented Open-source Project and the Latest Major Release (v5.0) of the Core Package}",
      journal = {\apj},
         year = 2022,
        month = aug,
       volume = {935},
       number = {2},
          eid = {167},
        pages = {167},
          doi = {10.3847/1538-4357/ac7c74},
archivePrefix = {arXiv},
       eprint = {2206.14220},
 primaryClass = {astro-ph.IM},
       adsurl = {https://ui.adsabs.harvard.edu/abs/2022ApJ...935..167A}
}

@ARTICLE{1979ApJ...230..204M,
author = {{Mouschovias}, T. Ch. and {Paleologou}, E.~V.},
title = "{The angular momentum problem and magnetic braking: an exact, time-dependent solution.}",
journal = {\apj},
year = 1979,
month = may,
volume = {230},
pages = {204-222},
doi = {10.1086/157077},
adsurl = {https://ui.adsabs.harvard.edu/abs/1979ApJ...230..204M}
}

@ARTICLE{2024ApJ...972L..27Y,
       author = {{Yen}, Hsi-Wei and {Lee}, Yueh-Ning},
        title = "{Protostellar Disk Formation Regimes: Angular Momentum Conservation versus Magnetic Braking}",
      journal = {\apjl},
         year = 2024,
        month = sep,
       volume = {972},
       number = {2},
          eid = {L27},
        pages = {L27},
          doi = {10.3847/2041-8213/ad7263},
archivePrefix = {arXiv},
       eprint = {2408.12101},
 primaryClass = {astro-ph.SR},
       adsurl = {https://ui.adsabs.harvard.edu/abs/2024ApJ...972L..27Y}
}

@ARTICLE{2025ApJ...988..125S,
       author = {{Shen}, Yu-Fu and {Xu}, Yan and {Wang}, Yi-Bo and {Huang}, Xiu-Lin and {Hu}, Xing-Xing and {Yuan}, Qi},
        title = "{The Angular Momentum of Stars Reflects the Relationship between Star-forming Environment and Galactic Evolution History}",
      journal = {\apj},
         year = 2025,
        month = jul,
       volume = {988},
       number = {1},
          eid = {125},
        pages = {125},
          doi = {10.3847/1538-4357/ade399},
archivePrefix = {arXiv},
       eprint = {2501.01035},
 primaryClass = {astro-ph.GA},
       adsurl = {https://ui.adsabs.harvard.edu/abs/2025ApJ...988..125S}
}

@ARTICLE{2021ApJ...920...83S,
       author = {{Sternberg}, Amiel and {Gurman}, Alon and {Bialy}, Shmuel},
        title = "{H I-to-H$_{2}$ Transitions in Dust-free Interstellar Gas}",
      journal = {\apj},
         year = 2021,
        month = oct,
       volume = {920},
       number = {2},
          eid = {83},
        pages = {83},
          doi = {10.3847/1538-4357/ac167b},
archivePrefix = {arXiv},
       eprint = {2105.01681},
 primaryClass = {astro-ph.GA},
       adsurl = {https://ui.adsabs.harvard.edu/abs/2021ApJ...920...83S}
}

@article{2022zndo...6053272R,
       author = {{Reback}, Jeff and {Jbrockmendel} and {McKinney}, Wes and {Van Den Bossche}, Joris and {Augspurger}, Tom and {Cloud}, Phillip and {Hawkins}, Simon and {Roeschke}, Matthew and {Gfyoung} and {Sinhrks} and {Klein}, Adam and {Hoefler}, Patrick and {Petersen}, Terji and {Tratner}, Jeff and {She}, Chang and {Ayd}, William and {Naveh}, Shahar and {Darbyshire}, JHM and {Garcia}, Marc and {Shadrach}, Richard and {Schendel}, Jeremy and {Hayden}, Andy and {Saxton}, Daniel and {Gorelli}, Marco Edward and {Li}, Fangchen and {Zeitlin}, Matthew and {Jancauskas}, Vytautas and {McMaster}, Ali and {Battiston}, Pietro and {Seabold}, Skipper},
        title = "{pandas-dev/pandas: Pandas 1.4.1}",
         year = 2022,
        month = feb,
          eid = {10.5281/zenodo.6053272},
          doi = {10.5281/zenodo.6053272},
      version = {v1.4.1},
    publisher = {Zenodo},
       adsurl = {https://ui.adsabs.harvard.edu/abs/2022zndo...6053272R}
}

@article{2018zndo...1218715V,
       author = {{Virtanen}, Pauli and {Gommers}, Ralf and {Burovski}, Evgeni and {Oliphant}, Travis E. and {Cournapeau}, David and {Weckesser}, Warren and {Alexbrc} and {Peterson}, Pearu and {Endolith} and {Van Der Walt}, Stefan and {Mayorov}, Nikolay and {Wilson}, Josh and {Laxalde}, Denis and {Brett}, Matthew and {Millman}, Jarrod and {Lars} and {Eric-Jones} and {Nelson}, Andrew and {Kern}, Robert and {Moore}, Eric and {Leslie}, Tim and {Perktold}, Josef and {Carey}, CJ and {Feng}, Yu and {Vanderplas}, Jake and {Haberland}, Matt and {Cowlicks} and {Larson}, Eric and {Polat}, Ilhan and {Reddy}, Tyler},
        title = "{scipy/scipy: Scipy 1.1.0rc1}",
         year = 2018,
        month = apr,
          eid = {10.5281/zenodo.1218715},
          doi = {10.5281/zenodo.1218715},
      version = {v1.1.0rc1},
    publisher = {Zenodo},
       adsurl = {https://ui.adsabs.harvard.edu/abs/2018zndo...1218715V}
}

@Article{harris2020array,
 author        = {Charles R. Harris and K. Jarrod Millman and St{\'{e}}fan J.
                 van der Walt and Ralf Gommers and Pauli Virtanen and David
                 Cournapeau and Eric Wieser and Julian Taylor and Sebastian
                 Berg and Nathaniel J. Smith and Robert Kern and Matti Picus
                 and Stephan Hoyer and Marten H. van Kerkwijk and Matthew
                 Brett and Allan Haldane and Jaime Fern{\'{a}}ndez del
                 R{\'{i}}o and Mark Wiebe and Pearu Peterson and Pierre
                 G{\'{e}}rard-Marchant and Kevin Sheppard and Tyler Reddy and
                 Warren Weckesser and Hameer Abbasi and Christoph Gohlke and
                 Travis E. Oliphant},
title          = {Array programming with {NumPy}},
 year          = {2020},
 month         = sep,
 journal       = {Nature},
 volume        = {585},
 number        = {7825},
 pages         = {357--362},
 doi           = {10.1038/s41586-020-2649-2},
 publisher     = {Springer Science and Business Media {LLC}},
 url           = {https://doi.org/10.1038/s41586-020-2649-2}
}

@ARTICLE{2022ApJ...925...78A,
       author = {{Arroyo-Ch{\'a}vez}, Griselda and {V{\'a}zquez-Semadeni}, Enrique},
        title = "{Evolution of the Angular Momentum during Gravitational Fragmentation of Molecular Clouds}",
      journal = {\apj},
         year = 2022,
        month = jan,
       volume = {925},
       number = {1},
          eid = {78},
        pages = {78},
          doi = {10.3847/1538-4357/ac3915},
archivePrefix = {arXiv},
       eprint = {2106.10381},
 primaryClass = {astro-ph.GA},
       adsurl = {https://ui.adsabs.harvard.edu/abs/2022ApJ...925...78A}
}

@ARTICLE{2000ApJ...543..822B,
       author = {{Burkert}, Andreas and {Bodenheimer}, Peter},
        title = "{Turbulent Molecular Cloud Cores: Rotational Properties}",
      journal = {\apj},
         year = 2000,
        month = nov,
       volume = {543},
       number = {2},
        pages = {822-830},
          doi = {10.1086/317122},
archivePrefix = {arXiv},
       eprint = {astro-ph/0006010},
 primaryClass = {astro-ph},
       adsurl = {https://ui.adsabs.harvard.edu/abs/2000ApJ...543..822B}
}

@ARTICLE{1993ApJ...406..528G,
       author = {{Goodman}, A.~A. and {Benson}, P.~J. and {Fuller}, G.~A. and {Myers}, P.~C.},
        title = "{Dense Cores in Dark Clouds. VIII. Velocity Gradients}",
      journal = {\apj},
         year = 1993,
        month = apr,
       volume = {406},
        pages = {528},
          doi = {10.1086/172465},
       adsurl = {https://ui.adsabs.harvard.edu/abs/1993ApJ...406..528G}
}

@ARTICLE{2011ApJ...732...78I,
       author = {{Imara}, Nia and {Blitz}, Leo},
        title = "{Angular Momentum in Giant Molecular Clouds. I. The Milky Way}",
      journal = {\apj},
         year = 2011,
        month = may,
       volume = {732},
       number = {2},
          eid = {78},
        pages = {78},
          doi = {10.1088/0004-637X/732/2/78},
archivePrefix = {arXiv},
       eprint = {1103.3741},
 primaryClass = {astro-ph.GA},
       adsurl = {https://ui.adsabs.harvard.edu/abs/2011ApJ...732...78I}
}

@ARTICLE{2011ApJ...732...79I,
       author = {{Imara}, Nia and {Bigiel}, Frank and {Blitz}, Leo},
        title = "{Angular Momentum in Giant Molecular Clouds. II. M33}",
      journal = {\apj},
         year = 2011,
        month = may,
       volume = {732},
       number = {2},
          eid = {79},
        pages = {79},
          doi = {10.1088/0004-637X/732/2/79},
archivePrefix = {arXiv},
       eprint = {1103.3702},
 primaryClass = {astro-ph.CO},
       adsurl = {https://ui.adsabs.harvard.edu/abs/2011ApJ...732...79I}
}

@ARTICLE{2022ApJ...931....9L,
       author = {{Lewis}, John Arban and {Lada}, Charles J. and {Dame}, T.~M.},
        title = "{Systematic Investigation of Dust and Gaseous CO in 12 Nearby Molecular Clouds}",
      journal = {\apj},
         year = 2022,
        month = may,
       volume = {931},
       number = {1},
          eid = {9},
        pages = {9},
          doi = {10.3847/1538-4357/ac5d58},
archivePrefix = {arXiv},
       eprint = {2203.07480},
 primaryClass = {astro-ph.GA},
       adsurl = {https://ui.adsabs.harvard.edu/abs/2022ApJ...931....9L}
}

@ARTICLE{2023ApJ...955...59P,
       author = {{Park}, Geumsook and {Koo}, Bon-Chul and {Kim}, Kee-Tae and {Elmegreen}, Bruce},
        title = "{Neutral Atomic and Molecular Clouds and Star Formation in the Outer Carina Arm}",
      journal = {\apj},
         year = 2023,
        month = sep,
       volume = {955},
       number = {1},
          eid = {59},
        pages = {59},
          doi = {10.3847/1538-4357/acebda},
archivePrefix = {arXiv},
       eprint = {2308.01577},
 primaryClass = {astro-ph.GA},
       adsurl = {https://ui.adsabs.harvard.edu/abs/2023ApJ...955...59P}
}

@ARTICLE{2011ApJ...740...45T,
       author = {{Tobin}, John J. and {Hartmann}, Lee and {Chiang}, Hsin-Fang and {Looney}, Leslie W. and {Bergin}, Edwin A. and {Chandler}, Claire J. and {Masqu{\'e}}, Josep M. and {Maret}, S{\'e}bastien and {Heitsch}, Fabian},
        title = "{Complex Structure in Class 0 Protostellar Envelopes. II. Kinematic Structure from Single-dish and Interferometric Molecular Line Mapping}",
      journal = {\apj},
         year = 2011,
        month = oct,
       volume = {740},
       number = {1},
          eid = {45},
        pages = {45},
          doi = {10.1088/0004-637X/740/1/45},
archivePrefix = {arXiv},
       eprint = {1107.4361},
 primaryClass = {astro-ph.SR},
       adsurl = {https://ui.adsabs.harvard.edu/abs/2011ApJ...740...45T}
}

@ARTICLE{2024ApJ...973L..27S,
       author = {{Sun}, Shenglan and {Wang}, Ke and {Liu}, Xunchuan and {Xu}, Fengwei},
        title = "{The Formation of Milky Way ``Bones'': Ubiquitous HI Narrow Self-absorption Associated with CO Emission}",
      journal = {\apjl},
         year = 2024,
        month = sep,
       volume = {973},
       number = {1},
          eid = {L27},
        pages = {L27},
          doi = {10.3847/2041-8213/ad77ce},
archivePrefix = {arXiv},
       eprint = {2409.01895},
 primaryClass = {astro-ph.GA},
       adsurl = {https://ui.adsabs.harvard.edu/abs/2024ApJ...973L..27S}
}

@ARTICLE{2018ApJ...865...34C,
       author = {{Chen}, Che-Yu and {Ostriker}, Eve C.},
        title = "{Geometry, Kinematics, and Magnetization of Simulated Prestellar Cores}",
      journal = {\apj},
         year = 2018,
        month = sep,
       volume = {865},
       number = {1},
          eid = {34},
        pages = {34},
          doi = {10.3847/1538-4357/aad905},
archivePrefix = {arXiv},
       eprint = {1808.02483},
 primaryClass = {astro-ph.SR},
       adsurl = {https://ui.adsabs.harvard.edu/abs/2018ApJ...865...34C}
}

@ARTICLE{2026arXiv260111830A,
       author = {{Arroyo-Chavez}, Griselda and {Vazquez-Semadeni}, Enrique and {Wurster}, James},
        title = "{On the role of gravity, turbulence, and the magnetic field in angular momentum transfer within molecular clouds}",
      journal = {arXiv e-prints},
         year = 2026,
        month = jan,
          eid = {arXiv:2601.11830},
        pages = {arXiv:2601.11830},
          doi = {10.48550/arXiv.2601.11830},
archivePrefix = {arXiv},
       eprint = {2601.11830},
 primaryClass = {astro-ph.GA},
       adsurl = {https://ui.adsabs.harvard.edu/abs/2026arXiv260111830A}
}

@article{joblib_development_team_2025_15496554,
  author       = {{The joblib developers}},
  title        = {joblib: running Python functions as pipeline jobs},
  month        = may,
  year         = 2025,
  publisher    = {Zenodo},
  version      = {latest},
  doi          = {10.5281/zenodo.15496554},
  url          = {https://doi.org/10.5281/zenodo.15496554}
}

@ARTICLE{2014ApJ...793..132S,
       author = {{Stanimirovi{\'c}}, Sne{\v{z}}ana and {Murray}, Claire E. and {Lee}, Min-Young and {Heiles}, Carl and {Miller}, Jesse},
        title = "{Cold and Warm Atomic Gas around the Perseus Molecular Cloud. I. Basic Properties}",
      journal = {\apj},
         year = 2014,
        month = oct,
       volume = {793},
       number = {2},
          eid = {132},
        pages = {132},
          doi = {10.1088/0004-637X/793/2/132},
archivePrefix = {arXiv},
       eprint = {1407.7778},
 primaryClass = {astro-ph.GA},
       adsurl = {https://ui.adsabs.harvard.edu/abs/2014ApJ...793..132S}
}

@ARTICLE{2003A&A...405..639P,
       author = {{Pirogov}, L. and {Zinchenko}, I. and {Caselli}, P. and {Johansson}, L.~E.~B. and {Myers}, P.~C.},
        title = "{N$_{2}$H$^{+}$(1-0) survey of massive molecular cloud cores}",
      journal = {\aap},
         year = 2003,
        month = jul,
       volume = {405},
        pages = {639-654},
          doi = {10.1051/0004-6361:20030659},
archivePrefix = {arXiv},
       eprint = {astro-ph/0304469},
 primaryClass = {astro-ph},
       adsurl = {https://ui.adsabs.harvard.edu/abs/2003A&A...405..639P}
}

@ARTICLE{2021MNRAS.505.4048L,
       author = {{Liu}, Lijie and {Bureau}, Martin and {Blitz}, Leo and {Davis}, Timothy A. and {Onishi}, Kyoko and {Smith}, Mark and {North}, Eve and {Iguchi}, Satoru},
        title = "{WISDOM Project - IX. Giant molecular clouds in the lenticular galaxy NGC 4429: effects of shear and tidal forces on clouds}",
      journal = {\mnras},
         year = 2021,
        month = aug,
       volume = {505},
       number = {3},
        pages = {4048-4085},
          doi = {10.1093/mnras/stab1537},
archivePrefix = {arXiv},
       eprint = {2106.04327},
 primaryClass = {astro-ph.GA},
       adsurl = {https://ui.adsabs.harvard.edu/abs/2021MNRAS.505.4048L}
}

@ARTICLE{2007ApJ...654..240R,
       author = {{Rosolowsky}, E.},
        title = "{Giant Molecular Clouds in M31. I. Molecular Cloud Properties}",
      journal = {\apj},
         year = 2007,
        month = jan,
       volume = {654},
       number = {1},
        pages = {240-251},
          doi = {10.1086/509249},
archivePrefix = {arXiv},
       eprint = {astro-ph/0609421},
 primaryClass = {astro-ph},
       adsurl = {https://ui.adsabs.harvard.edu/abs/2007ApJ...654..240R}
}

@ARTICLE{2018A&A...617A..27P,
       author = {{Punanova}, A. and {Caselli}, P. and {Pineda}, J.~E. and {Pon}, A. and {Tafalla}, M. and {Hacar}, A. and {Bizzocchi}, L.},
        title = "{Kinematics of dense gas in the L1495 filament}",
      journal = {\aap},
         year = 2018,
        month = sep,
       volume = {617},
          eid = {A27},
        pages = {A27},
          doi = {10.1051/0004-6361/201731159},
archivePrefix = {arXiv},
       eprint = {1806.03354},
 primaryClass = {astro-ph.GA},
       adsurl = {https://ui.adsabs.harvard.edu/abs/2018A&A...617A..27P}
}

@ARTICLE{2015ApJ...799..193Y,
       author = {{Yen}, Hsi-Wei and {Koch}, Patrick M. and {Takakuwa}, Shigehisa and {Ho}, Paul T.~P. and {Ohashi}, Nagayoshi and {Tang}, Ya-Wen},
        title = "{Observations of Infalling and Rotational Motions on a 1000 AU Scale around 17 Class 0 and 0/I Protostars: Hints of Disk Growth and Magnetic Braking?}",
      journal = {\apj},
         year = 2015,
        month = feb,
       volume = {799},
       number = {2},
          eid = {193},
        pages = {193},
          doi = {10.1088/0004-637X/799/2/193},
archivePrefix = {arXiv},
       eprint = {1412.1916},
 primaryClass = {astro-ph.SR},
       adsurl = {https://ui.adsabs.harvard.edu/abs/2015ApJ...799..193Y}
}

@ARTICLE{2019MNRAS.490..527C,
       author = {{Chen}, Che-Yu and {Storm}, Shaye and {Li}, Zhi-Yun and {Mundy}, Lee G. and {Frayer}, David and {Li}, Jialu and {Church}, Sarah and {Friesen}, Rachel and {Harris}, Andrew I. and {Looney}, Leslie W. and {Offner}, Stella and {Ostriker}, Eve C. and {Pineda}, Jaime E. and {Tobin}, John and {Chen}, Hope H.-H.},
        title = "{Investigating the complex velocity structures within dense molecular cloud cores with GBT-Argus}",
      journal = {\mnras},
         year = 2019,
        month = nov,
       volume = {490},
       number = {1},
        pages = {527-539},
          doi = {10.1093/mnras/stz2633},
archivePrefix = {arXiv},
       eprint = {1909.07997},
 primaryClass = {astro-ph.GA},
       adsurl = {https://ui.adsabs.harvard.edu/abs/2019MNRAS.490..527C}
}

@ARTICLE{2023MNRAS.525..364P,
       author = {{Pandhi}, A. and {Friesen}, R.~K. and {Fissel}, L. and {Pineda}, J.~E. and {Caselli}, P. and {Chen}, M.~C.-Y. and {Di Francesco}, J. and {Ginsburg}, A. and {Kirk}, H. and {Myers}, P.~C. and {Offner}, S.~S.~R. and {Punanova}, A. and {Quan}, F. and {Redaelli}, E. and {Rosolowsky}, E. and {Scibelli}, S. and {Seo}, Y.~M. and {Shirley}, Y.},
        title = "{Alignment of dense molecular core morphology and velocity gradients with ambient magnetic fields}",
      journal = {\mnras},
         year = 2023,
        month = oct,
       volume = {525},
       number = {1},
        pages = {364-392},
          doi = {10.1093/mnras/stad2283},
archivePrefix = {arXiv},
       eprint = {2307.13022},
 primaryClass = {astro-ph.GA},
       adsurl = {https://ui.adsabs.harvard.edu/abs/2023MNRAS.525..364P}
}

@ARTICLE{1999A&AS..134..241P,
       author = {{Phillips}, J.~P.},
        title = "{Rotation in molecular clouds}",
      journal = {\aaps},
         year = 1999,
        month = jan,
       volume = {134},
        pages = {241-254},
          doi = {10.1051/aas:1999137},
       adsurl = {https://ui.adsabs.harvard.edu/abs/1999A&AS..134..241P}
}

@article{1998yCat..41340241P,
       author = {{Phillips}, J.~P.},
        title = "{VizieR Online Data Catalog: Rotation in molecular clouds (Phillips 1999)}",
 howpublished = {VizieR On-line Data Catalog: J/A+AS/134/241. Originally published in: 1999A\&AS..134..241P},
         year = 1998,
        month = aug,
          eid = {J/A+AS/134/241},
       adsurl = {https://ui.adsabs.harvard.edu/abs/1998yCat..41340241P}
}

@ARTICLE{2002ApJ...572..238C,
       author = {{Caselli}, Paola and {Benson}, Priscilla J. and {Myers}, Philip C. and {Tafalla}, Mario},
        title = "{Dense Cores in Dark Clouds. XIV. N$_{2}$H$^{+}$ (1-0) Maps of Dense Cloud Cores}",
      journal = {\apj},
         year = 2002,
        month = jun,
       volume = {572},
       number = {1},
        pages = {238-263},
          doi = {10.1086/340195},
archivePrefix = {arXiv},
       eprint = {astro-ph/0202173},
 primaryClass = {astro-ph},
       adsurl = {https://ui.adsabs.harvard.edu/abs/2002ApJ...572..238C}
}

@ARTICLE{1981MNRAS.194..809L,
       author = {{Larson}, R.~B.},
        title = "{Turbulence and star formation in molecular clouds.}",
      journal = {\mnras},
         year = 1981,
        month = mar,
       volume = {194},
        pages = {809-826},
          doi = {10.1093/mnras/194.4.809},
       adsurl = {https://ui.adsabs.harvard.edu/abs/1981MNRAS.194..809L}
}

@INPROCEEDINGS{1993prpl.conf..125B,
       author = {{Blitz}, Leo},
        title = "{Giant Molecular Clouds}",
    booktitle = {Protostars and Planets III},
         year = 1993,
       editor = {{Levy}, Eugene H. and {Lunine}, Jonathan I.},
        month = jan,
        pages = {125},
       adsurl = {https://ui.adsabs.harvard.edu/abs/1993prpl.conf..125B}
}

@ARTICLE{2024MNRAS.532.3480D,
       author = {{Dharmawardena}, T.~E. and {Bailer-Jones}, C.~A.~L. and {Fouesneau}, M. and {Foreman-Mackey}, D. and {Coronica}, P. and {Colnaghi}, T. and {M{\"u}ller}, T. and {Wilson}, A.~G.},
        title = "{All-sky three-dimensional dust density and extinction Maps of the Milky Way out to 2.8 kpc}",
      journal = {\mnras},
         year = 2024,
        month = aug,
       volume = {532},
       number = {3},
        pages = {3480-3498},
          doi = {10.1093/mnras/stae1474},
archivePrefix = {arXiv},
       eprint = {2406.06740},
 primaryClass = {astro-ph.GA},
       adsurl = {https://ui.adsabs.harvard.edu/abs/2024MNRAS.532.3480D}
}

@ARTICLE{2007ApJ...669.1058C,
       author = {{Chen}, Xuepeng and {Launhardt}, Ralf and {Henning}, Thomas},
        title = "{OVRO N$_{2}$H$^{+}$ Observations of Class 0 Protostars: Constraints on the Formation of Binary Stars}",
      journal = {\apj},
         year = 2007,
        month = nov,
       volume = {669},
       number = {2},
        pages = {1058-1071},
          doi = {10.1086/521868},
archivePrefix = {arXiv},
       eprint = {0707.4419},
 primaryClass = {astro-ph},
       adsurl = {https://ui.adsabs.harvard.edu/abs/2007ApJ...669.1058C}
}

@ARTICLE{2019ApJ...886..119C,
       author = {{Chen}, Hope How-Huan and {Pineda}, Jaime E. and {Offner}, Stella S.~R. and {Goodman}, Alyssa A. and {Burkert}, Andreas and {Friesen}, Rachel K. and {Rosolowsky}, Erik and {Scibelli}, Samantha and {Shirley}, Yancy},
        title = "{Droplets. II. Internal Velocity Structures and Potential Rotational Motions in Pressure-dominated Coherent Structures}",
      journal = {\apj},
         year = 2019,
        month = dec,
       volume = {886},
       number = {2},
          eid = {119},
        pages = {119},
          doi = {10.3847/1538-4357/ab4ce9},
archivePrefix = {arXiv},
       eprint = {1908.04367},
 primaryClass = {astro-ph.GA},
       adsurl = {https://ui.adsabs.harvard.edu/abs/2019ApJ...886..119C}
}

@ARTICLE{2016PASJ...68...24T,
       author = {{Tatematsu}, Ken'ichi and {Ohashi}, Satoshi and {Sanhueza}, Patricio and {Nguyen Luong}, Quang and {Umemoto}, Tomofumi and {Mizuno}, Norikazu},
        title = "{Angular momentum of the N$_{2}$H$^{+}$ cores in the Orion A cloud}",
      journal = {\pasj},
         year = 2016,
        month = apr,
       volume = {68},
       number = {2},
          eid = {24},
        pages = {24},
          doi = {10.1093/pasj/psw002},
archivePrefix = {arXiv},
       eprint = {1601.00362},
 primaryClass = {astro-ph.GA},
       adsurl = {https://ui.adsabs.harvard.edu/abs/2016PASJ...68...24T}
}

@ARTICLE{2024A&A...685A..82E,
       author = {{Edenhofer}, Gordian and {Zucker}, Catherine and {Frank}, Philipp and {Saydjari}, Andrew K. and {Speagle}, Joshua S. and {Finkbeiner}, Douglas and {En{\ss}lin}, Torsten A.},
        title = "{A parsec-scale Galactic 3D dust map out to 1.25 kpc from the Sun}",
      journal = {\aap},
         year = 2024,
        month = may,
       volume = {685},
          eid = {A82},
        pages = {A82},
          doi = {10.1051/0004-6361/202347628},
archivePrefix = {arXiv},
       eprint = {2308.01295},
 primaryClass = {astro-ph.GA},
       adsurl = {https://ui.adsabs.harvard.edu/abs/2024A&A...685A..82E}
}

@ARTICLE{2020A&A...639A.138L,
       author = {{Leike}, R.~H. and {Glatzle}, M. and {En{\ss}lin}, T.~A.},
        title = "{Resolving nearby dust clouds}",
      journal = {\aap},
         year = 2020,
        month = jul,
       volume = {639},
          eid = {A138},
        pages = {A138},
          doi = {10.1051/0004-6361/202038169},
archivePrefix = {arXiv},
       eprint = {2004.06732},
 primaryClass = {astro-ph.GA},
       adsurl = {https://ui.adsabs.harvard.edu/abs/2020A&A...639A.138L}
}

@ARTICLE{2011arXiv1101.1499D,
       author = {{Dame}, T.~M.},
        title = "{Optimization of Moment Masking for CO Spectral Line Surveys}",
      journal = {arXiv e-prints},
         year = 2011,
        month = jan,
          eid = {arXiv:1101.1499},
        pages = {arXiv:1101.1499},
          doi = {10.48550/arXiv.1101.1499},
archivePrefix = {arXiv},
       eprint = {1101.1499},
 primaryClass = {astro-ph.IM},
       adsurl = {https://ui.adsabs.harvard.edu/abs/2011arXiv1101.1499D}
}

@ARTICLE{2001ApJ...547..792D,
       author = {{Dame}, T.~M. and {Hartmann}, Dap and {Thaddeus}, P.},
        title = "{The Milky Way in Molecular Clouds: A New Complete CO Survey}",
      journal = {\apj},
         year = 2001,
        month = feb,
       volume = {547},
       number = {2},
        pages = {792-813},
          doi = {10.1086/318388},
archivePrefix = {arXiv},
       eprint = {astro-ph/0009217},
 primaryClass = {astro-ph},
       adsurl = {https://ui.adsabs.harvard.edu/abs/2001ApJ...547..792D}
}

@ARTICLE{2016A&A...594A.116H,
       author = {{HI4PI Collaboration} and {Ben Bekhti}, N. and {Fl{\"o}er}, L. and {Keller}, R. and {Kerp}, J. and {Lenz}, D. and {Winkel}, B. and {Bailin}, J. and {Calabretta}, M.~R. and {Dedes}, L. and {Ford}, H.~A. and {Gibson}, B.~K. and {Haud}, U. and {Janowiecki}, S. and {Kalberla}, P.~M.~W. and {Lockman}, F.~J. and {McClure-Griffiths}, N.~M. and {Murphy}, T. and {Nakanishi}, H. and {Pisano}, D.~J. and {Staveley-Smith}, L.},
        title = "{HI4PI: A full-sky H I survey based on EBHIS and GASS}",
      journal = {\aap},
         year = 2016,
        month = oct,
       volume = {594},
          eid = {A116},
        pages = {A116},
          doi = {10.1051/0004-6361/201629178},
archivePrefix = {arXiv},
       eprint = {1610.06175},
 primaryClass = {astro-ph.GA},
       adsurl = {https://ui.adsabs.harvard.edu/abs/2016A&A...594A.116H}
}

@ARTICLE{2019ApJ...879..125Z,
       author = {{Zucker}, Catherine and {Speagle}, Joshua S. and {Schlafly}, Edward F. and {Green}, Gregory M. and {Finkbeiner}, Douglas P. and {Goodman}, Alyssa A. and {Alves}, Jo{\~a}o},
        title = "{A Large Catalog of Accurate Distances to Local Molecular Clouds: The Gaia DR2 Edition}",
      journal = {\apj},
         year = 2019,
        month = jul,
       volume = {879},
       number = {2},
          eid = {125},
        pages = {125},
          doi = {10.3847/1538-4357/ab2388},
archivePrefix = {arXiv},
       eprint = {1902.01425},
 primaryClass = {astro-ph.GA},
       adsurl = {https://ui.adsabs.harvard.edu/abs/2019ApJ...879..125Z}
}

@ARTICLE{2003ApJ...587..278W,
       author = {{Wolfire}, Mark G. and {McKee}, Christopher F. and {Hollenbach}, David and {Tielens}, A.~G.~G.~M.},
        title = "{Neutral Atomic Phases of the Interstellar Medium in the Galaxy}",
      journal = {\apj},
         year = 2003,
        month = apr,
       volume = {587},
       number = {1},
        pages = {278-311},
          doi = {10.1086/368016},
archivePrefix = {arXiv},
       eprint = {astro-ph/0207098},
 primaryClass = {astro-ph},
       adsurl = {https://ui.adsabs.harvard.edu/abs/2003ApJ...587..278W}
}

@misc{https://doi.org/10.26093/cds/vizier.35940116,
doi = {10.26093/CDS/VIZIER.35940116},
url = {https://cdsarc.cds.unistra.fr/viz-bin/cat/J/A+A/594/A116},
author = {{HI4PI Collaboration} and Ben Bekhti, N. and Floeer, L. and Keller, R. and Kerp, J. and Lenz, D. and Winkel, B. and Bailin, J. and Calabretta, M.R. and Dedes, L. and Ford, H.A. and Gibson, B.K. and Haud, U. and Janowiecki, S. and Kalberla P. M., W. and Lockman, F.J. and McClure-Griffiths, N.M. and Murphy, T. and Nakanishi, H. and Pisano, D.J. and Staveley-Smith, L.},
title = {HI4PI spectra and column density maps},
publisher = {Centre de Donnees Strasbourg (CDS)},
year = {2016},
copyright = {Refer to CDS usage}
}

@ARTICLE{2009ApJ...693..216K,
       author = {{Krumholz}, Mark R. and {McKee}, Christopher F. and {Tumlinson}, Jason},
        title = "{The Atomic-to-Molecular Transition in Galaxies. II: H I and H$_{2}$ Column Densities}",
      journal = {\apj},
         year = 2009,
        month = mar,
       volume = {693},
       number = {1},
        pages = {216-235},
          doi = {10.1088/0004-637X/693/1/216},
archivePrefix = {arXiv},
       eprint = {0811.0004},
 primaryClass = {astro-ph},
       adsurl = {https://ui.adsabs.harvard.edu/abs/2009ApJ...693..216K}
}

@ARTICLE{2018MNRAS.480L.126S,
       author = {{Saha}, Preetha and {Roy}, Nirupam and {Bhattacharya}, Mukul},
        title = "{On estimating the atomic hydrogen column density from the H I 21 cm emission spectra}",
      journal = {\mnras},
         year = 2018,
        month = oct,
       volume = {480},
       number = {1},
        pages = {L126-L130},
          doi = {10.1093/mnrasl/sly139},
archivePrefix = {arXiv},
       eprint = {1807.11497},
 primaryClass = {astro-ph.GA},
       adsurl = {https://ui.adsabs.harvard.edu/abs/2018MNRAS.480L.126S}
}

@ARTICLE{2005A&A...440..775K,
       author = {{Kalberla}, P.~M.~W. and {Burton}, W.~B. and {Hartmann}, Dap and {Arnal}, E.~M. and {Bajaja}, E. and {Morras}, R. and {P{\"o}ppel}, W.~G.~L.},
        title = "{The Leiden/Argentine/Bonn (LAB) Survey of Galactic HI. Final data release of the combined LDS and IAR surveys with improved stray-radiation corrections}",
      journal = {\aap},
         year = 2005,
        month = sep,
       volume = {440},
       number = {2},
        pages = {775-782},
          doi = {10.1051/0004-6361:20041864},
archivePrefix = {arXiv},
       eprint = {astro-ph/0504140},
 primaryClass = {astro-ph},
       adsurl = {https://ui.adsabs.harvard.edu/abs/2005A&A...440..775K}
}

@ARTICLE{2003ApJ...599..258R,
       author = {{Rosolowsky}, E. and {Engargiola}, G. and {Plambeck}, R. and {Blitz}, L.},
        title = "{Giant Molecular Clouds in M33. II. High-Resolution Observations}",
      journal = {\apj},
         year = 2003,
        month = dec,
       volume = {599},
       number = {1},
        pages = {258-274},
          doi = {10.1086/379166},
archivePrefix = {arXiv},
       eprint = {astro-ph/0307322},
 primaryClass = {astro-ph},
       adsurl = {https://ui.adsabs.harvard.edu/abs/2003ApJ...599..258R}
}

@ARTICLE{2016ApJ...829..102I,
       author = {{Imara}, Nia and {Burkhart}, Blakesley},
        title = "{The H I Probability Distribution Function and the Atomic-to-molecular Transition in Molecular Clouds}",
      journal = {\apj},
         year = 2016,
        month = oct,
       volume = {829},
       number = {2},
          eid = {102},
        pages = {102},
          doi = {10.3847/0004-637X/829/2/102},
archivePrefix = {arXiv},
       eprint = {1609.04817},
 primaryClass = {astro-ph.GA},
       adsurl = {https://ui.adsabs.harvard.edu/abs/2016ApJ...829..102I}
}

@ARTICLE{1977ApJ...215..521K,
       author = {{Kutner}, M.~L. and {Tucker}, K.~D. and {Chin}, G. and {Thaddeus}, P.},
        title = "{The molecular complexes in Orion.}",
      journal = {\apj},
         year = 1977,
        month = jul,
       volume = {215},
        pages = {521-528},
          doi = {10.1086/155384},
       adsurl = {https://ui.adsabs.harvard.edu/abs/1977ApJ...215..521K}
}

@article{Xu__2020,
   title={Rotation of Two Micron All Sky Survey Clumps in Molecular Clouds},
   volume={898},
   ISSN={1538-4357},
   url={http://dx.doi.org/10.3847/1538-4357/ab9a45},
   DOI={10.3847/1538-4357/ab9a45},
   number={2},
   journal={The Astrophysical Journal},
   publisher={American Astronomical Society},
   author={Xu, Xuefang and Li, Di and Dai, Y. Sophia and Goldsmith, Paul F. and Fuller, Gary A.},
   year={2020},
   month=jul, pages={122} }

@ARTICLE{2024arXiv240810406V,
       author = {{V{\'a}zquez-Semadeni}, Enrique and {Palau}, Aina and {G{\'o}mez}, Gilberto C. and {Arroyo-Ch{\'a}vez}, Griselda and {Alig}, Christian and {Ballesteros-Paredes}, Javier and {Camacho}, Vianey and {Gonz{\'a}lez-Samaniego}, Alejandro and {Burkert}, Andreas},
        title = "{The GT and GHC models for molecular clouds compared. Differences, similarities, and myths}",
      journal = {arXiv e-prints},
         year = 2024,
        month = aug,
          eid = {arXiv:2408.10406},
        pages = {arXiv:2408.10406},
          doi = {10.48550/arXiv.2408.10406},
archivePrefix = {arXiv},
       eprint = {2408.10406},
 primaryClass = {astro-ph.GA},
       adsurl = {https://ui.adsabs.harvard.edu/abs/2024arXiv240810406V}
}

@ARTICLE{2020A&A...633A..17B,
       author = {{Braine}, J. and {Hughes}, A. and {Rosolowsky}, E. and {Gratier}, P. and {Colombo}, D. and {Meidt}, S. and {Schinnerer}, E.},
        title = "{Rotation of molecular clouds in M 51}",
      journal = {\aap},
         year = 2020,
        month = jan,
       volume = {633},
          eid = {A17},
        pages = {A17},
          doi = {10.1051/0004-6361/201834613},
archivePrefix = {arXiv},
       eprint = {1911.08977},
 primaryClass = {astro-ph.GA},
       adsurl = {https://ui.adsabs.harvard.edu/abs/2020A&A...633A..17B}
}

@ARTICLE{2018A&A...612A..51B,
       author = {{Braine}, J. and {Rosolowsky}, E. and {Gratier}, P. and {Corbelli}, E. and {Schuster}, K. -F.},
        title = "{Properties and rotation of molecular clouds in M 33}",
      journal = {\aap},
         year = 2018,
        month = apr,
       volume = {612},
          eid = {A51},
        pages = {A51},
          doi = {10.1051/0004-6361/201732405},
archivePrefix = {arXiv},
       eprint = {1801.04171},
 primaryClass = {astro-ph.GA},
       adsurl = {https://ui.adsabs.harvard.edu/abs/2018A&A...612A..51B}
}

@article{Jeffreson_2020,
   title={The role of galactic dynamics in shaping the physical properties of giant molecular clouds in Milky Way-like galaxies},
   volume={498},
   ISSN={1365-2966},
   url={http://dx.doi.org/10.1093/mnras/staa2127},
   DOI={10.1093/mnras/staa2127},
   number={1},
   journal={Monthly Notices of the Royal Astronomical Society},
   publisher={Oxford University Press (OUP)},
   author={Jeffreson, Sarah M R and Kruijssen, J M Diederik and Keller, Benjamin W and Chevance, Mélanie and Glover, Simon C O},
   year={2020},
   month=jul, pages={385–429} }
\bibliographystyle{aasjournal}


\appendix
\counterwithin{figure}{section}
\section{Velocity Spectra}\label{sec:appndx_specs}
We calculate the one-dimensional velocity spectrum of each molecular cloud by taking the average brightness temperature of $^{12}\mathrm{CO}$ emission within a subcube, extracted from \cite{2001ApJ...547..792D} according to the latitude and longitude bounds given in Table \ref{tab:intro}. We sum emission in the range $v_\mathrm{H_2} \in \pm 50\ \mathrm{km\ s^{-1}}$. The HI envelope spectra are calculated from subcubes extracted from \cite{https://doi.org/10.26093/cds/vizier.35940116}. We define the center of the HI envelope to coincide with the center of the molecular cloud, and its latitude and longitude bounds define a square of side length of $2R_\mathrm{HI}$. To convert from linear to angular units, we assume that the HI envelope is located at the same distance as the molecular cloud with which it is associated. We fit the 21-cm spectrum with a \texttt{Gaussian1D} model from \texttt{astropy.modeling}, using the Trust Region Reflective algorithm in \texttt{astropy.fitting} to obtain the linecenter $v_\mathrm{HI}$ and linewidth  $\sigma_\mathrm{HI}$.
\begin{figure}[h!]
\includegraphics[width=\linewidth]{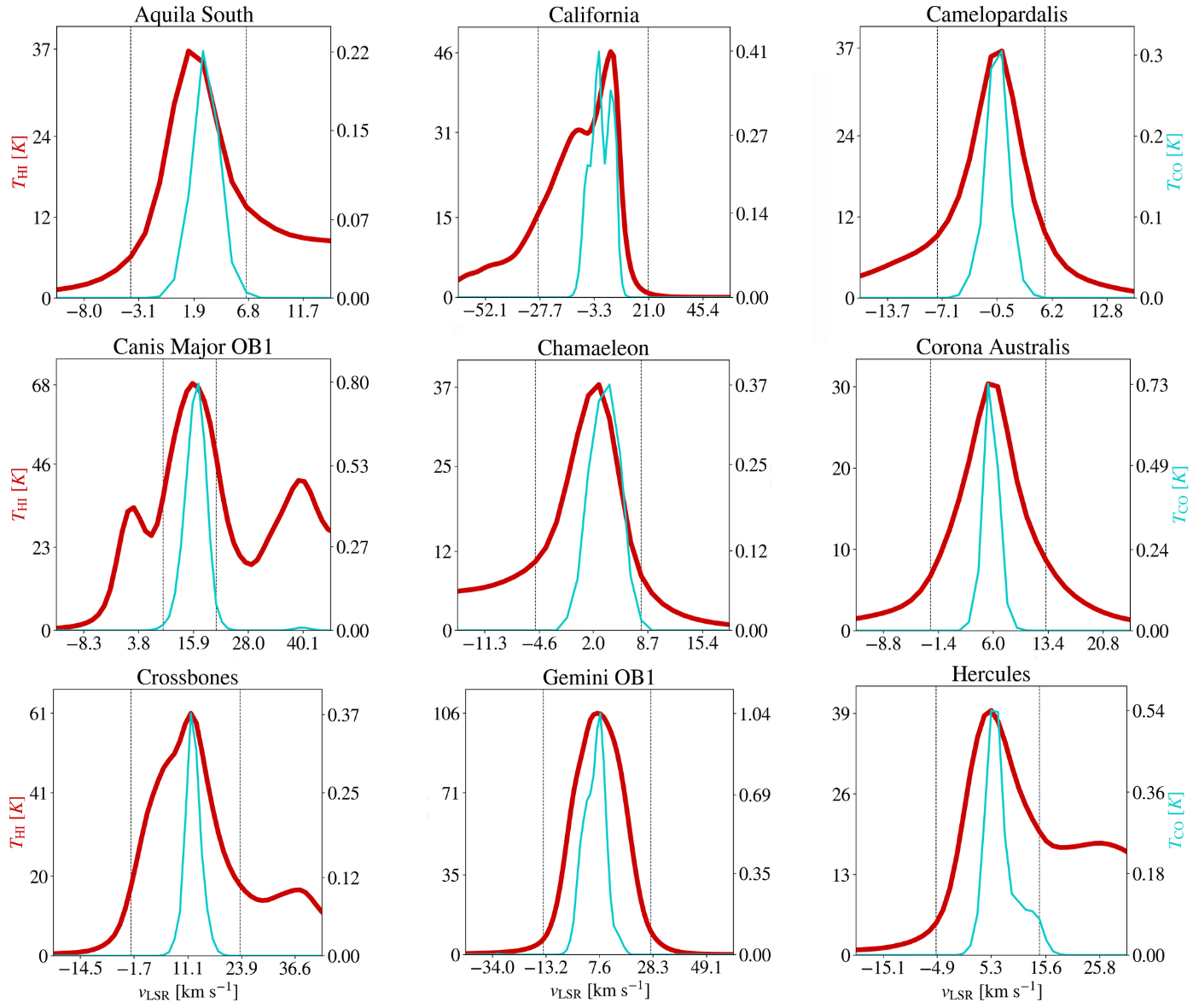}
\caption{Velocity spectra of the molecular clouds (blue) and their associated HI envelopes (red). The vertical lines indicate velocity bounds used to define the HI envelope. }
\label{fig:appndx_spec1}
\end{figure}

\begin{figure}[ht!]
\includegraphics[width=\linewidth]{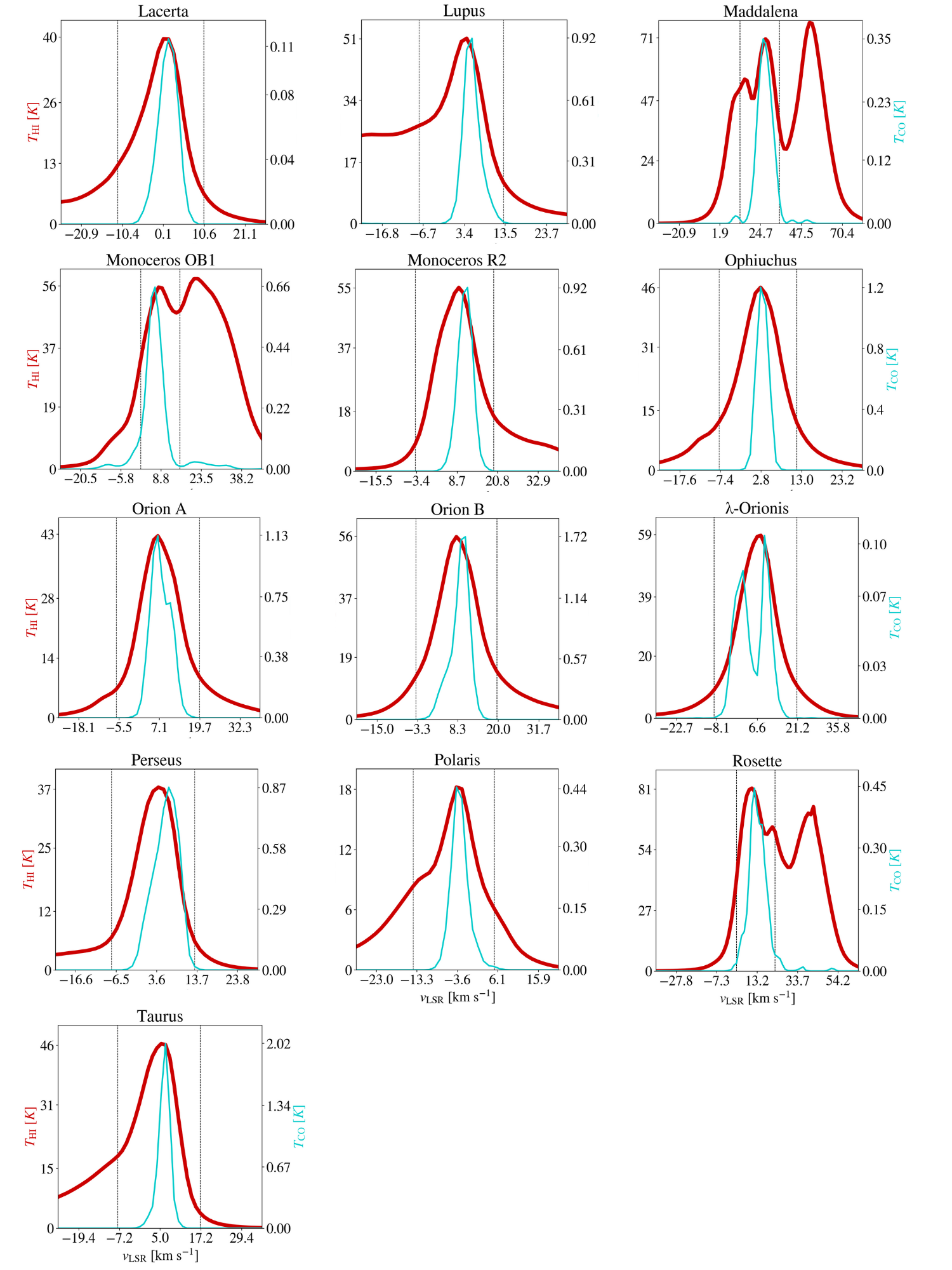}
\caption{Same as Figure \ref{fig:appndx_spec1}}
\label{fig:appndx_spec2}
\end{figure}

\newpage
\section{Surface density and velocity field maps}\label{sec:appndx_maps}
The surface density maps of the HI envelopes are presented below, along with the velocity field maps of both populations. In all the systems, the peak $\mathrm{H_2}$ surface density is much larger than the $\Sigma_\mathrm{HI,peak}$. For the Orion A and B clouds, the molecular gas is $>20\times$ denser than the surrounding HI. 

\begin{figure}[h!]
\includegraphics[width=\linewidth]{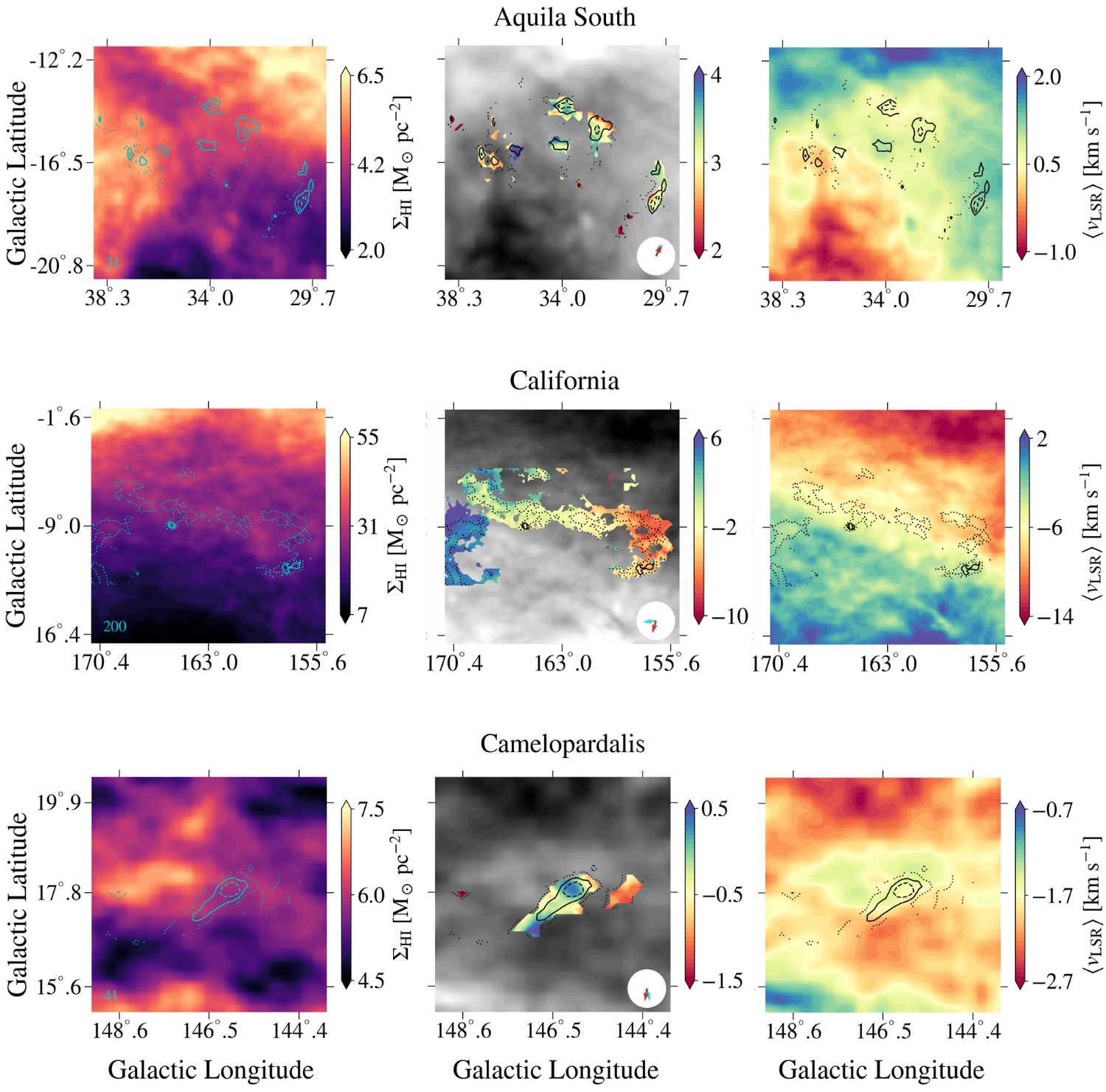}
\caption{\textbf{Surface} density map of the HI envelope around the Perseus molecular cloud (black outline). The overlaid blue contours are at $(0.25,\ 0.5,\ 0.75) \times \Sigma_\mathrm{H_2,max}$, which is indicated in lower left corner. \textit{Middle}: Velocity field maps of the Perseus molecular cloud and (\textit{right}) HI envelope, derived from the first moments of $^{12}\mathrm{CO}$ and 21-cm emission at each pixel. The blue (red) arrow in the lower right corner of the middle panel indicates the velocity gradient direction $\theta_\mathrm{H_2}$ ($\theta_\mathrm{HI}$).}
\label{fig:appndx_maps1}
\end{figure}

\begin{figure}[ht!]
\includegraphics[width=\linewidth]{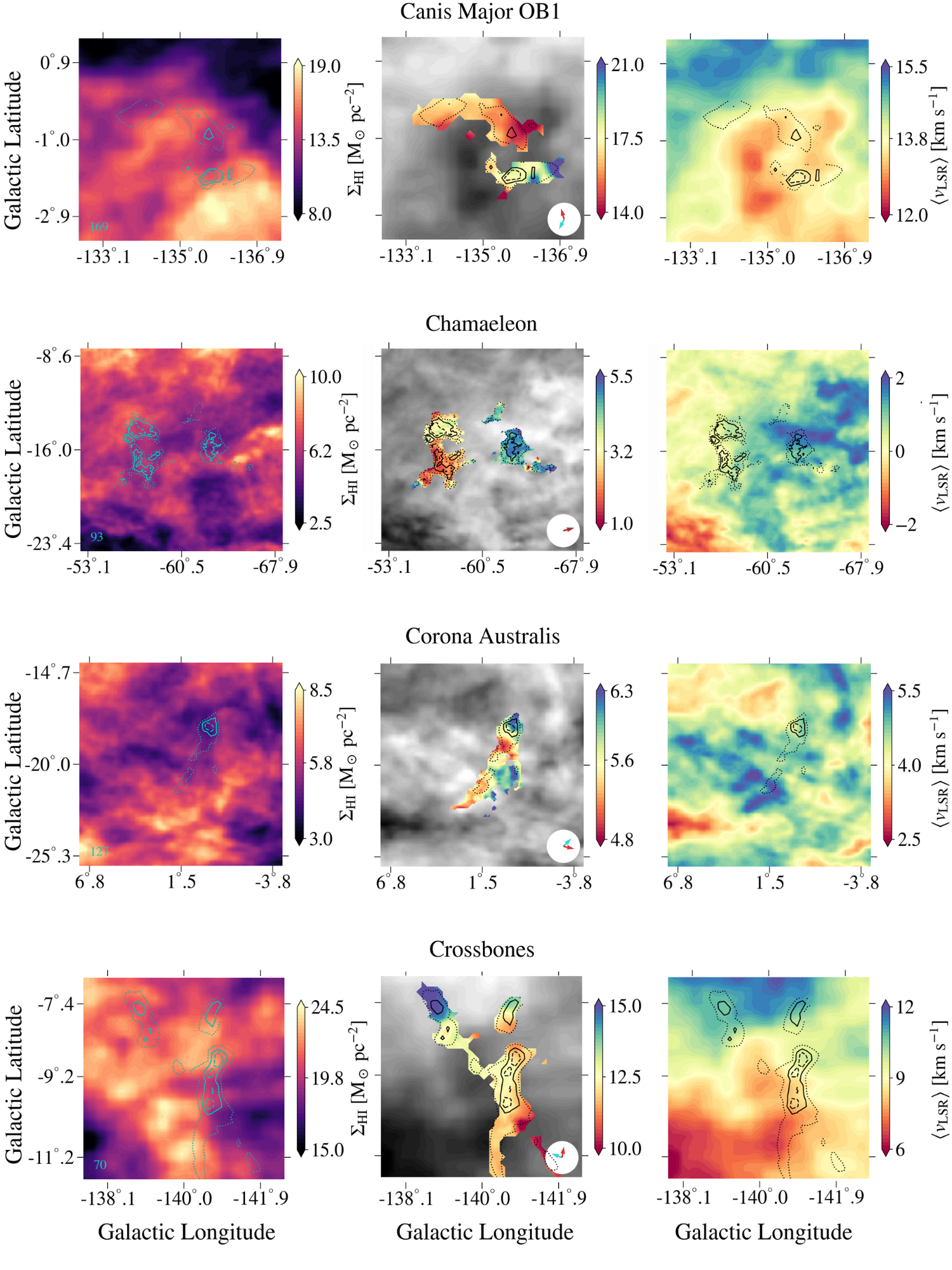}
\caption{Same as Figure \ref{fig:appndx_maps1}}
\label{fig:appndx_maps2}
\end{figure}

\begin{figure}[ht!]
\includegraphics[width=\linewidth]{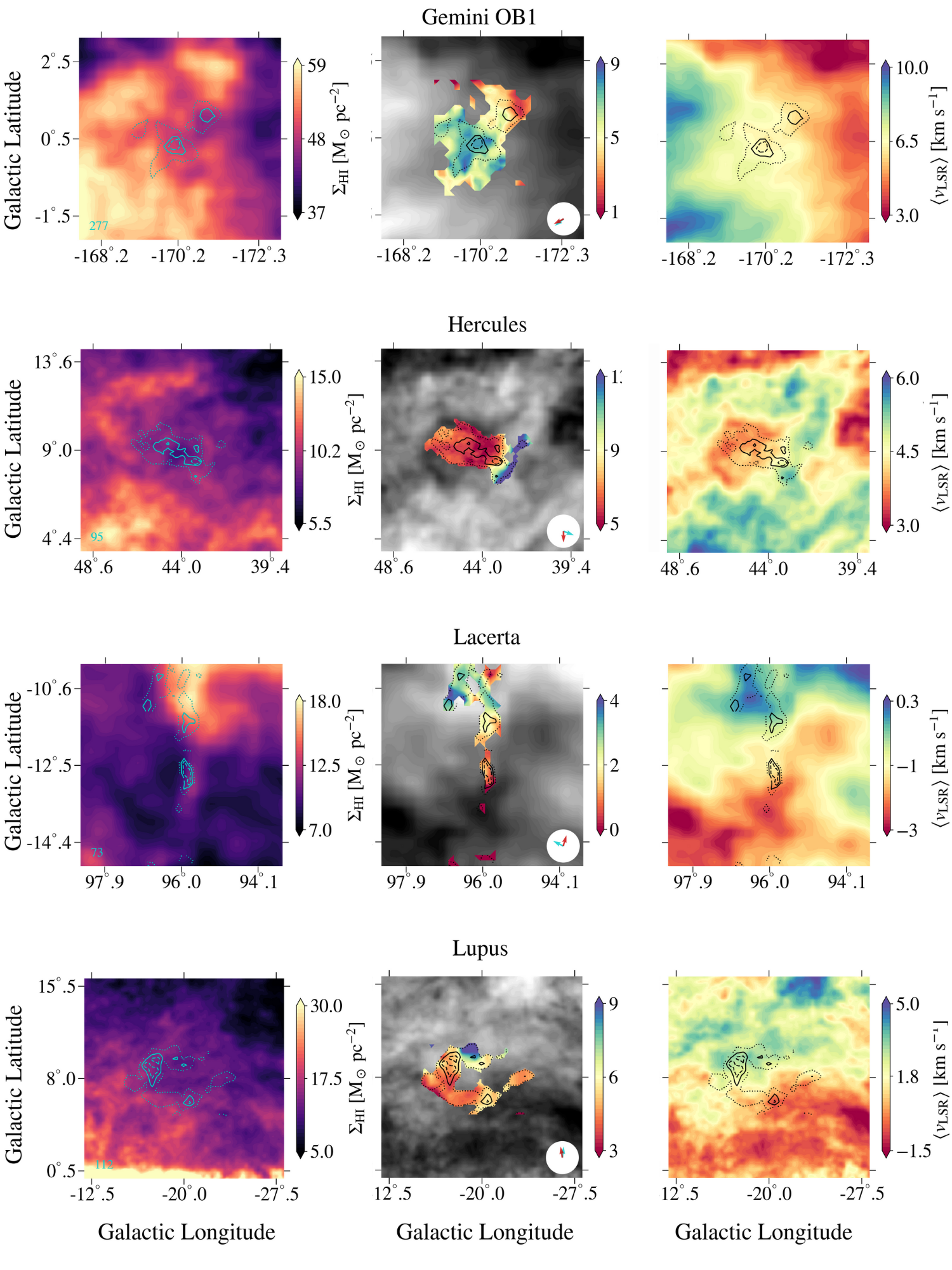}
\caption{Same as Figure \ref{fig:appndx_maps1}}
\label{fig:appndx_maps3}
\end{figure}

\begin{figure}[ht!]
\includegraphics[width=\linewidth]{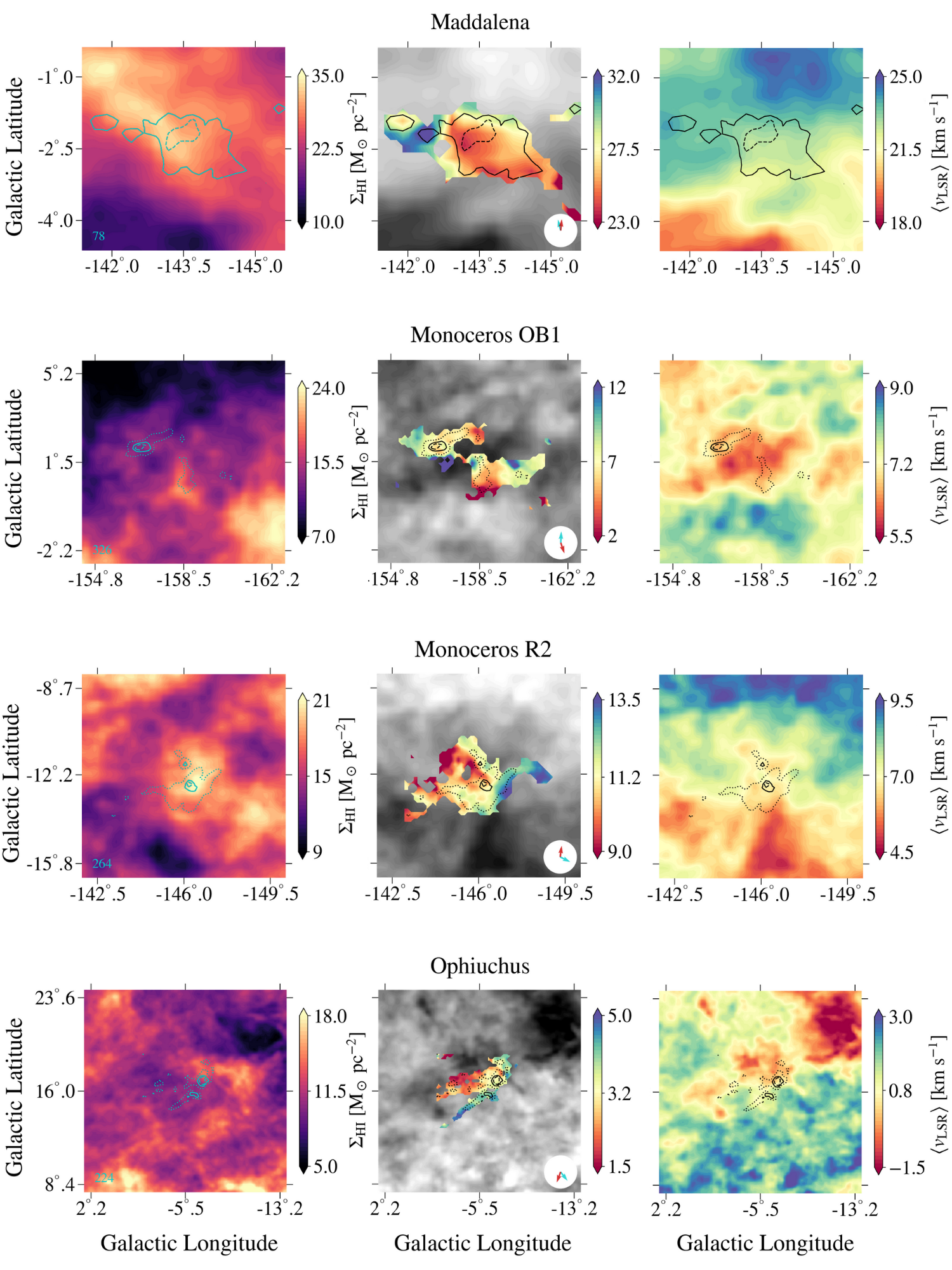}
\caption{Same as Figure \ref{fig:appndx_maps1}}
\label{fig:appndx_maps4}
\end{figure}

\begin{figure}[ht!]
\includegraphics[width=\linewidth]{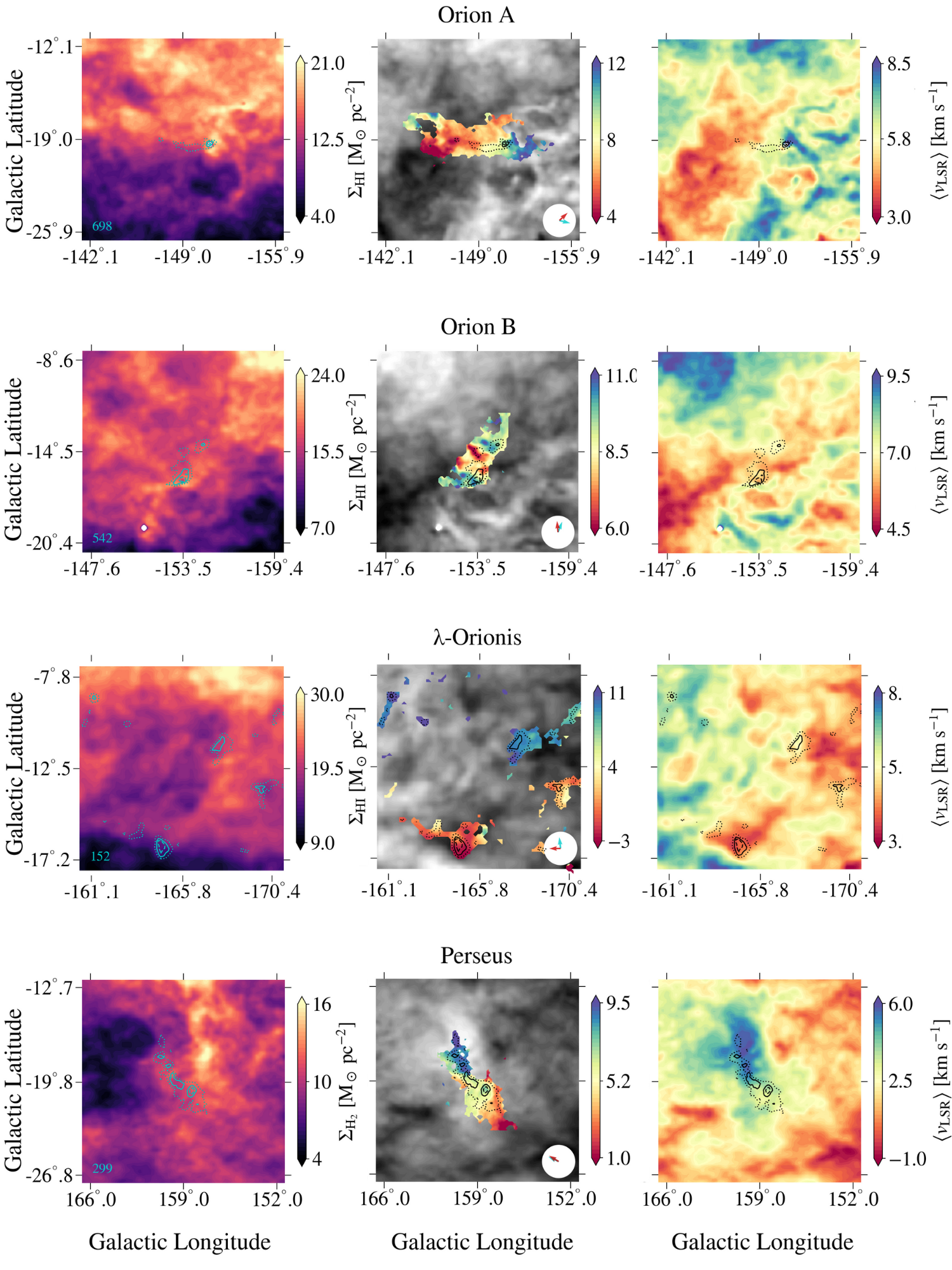}
\caption{Same as Figure \ref{fig:appndx_maps1}}
\label{fig:appndx_maps5}
\end{figure}

\begin{figure}[ht!]
\includegraphics[width=\linewidth]{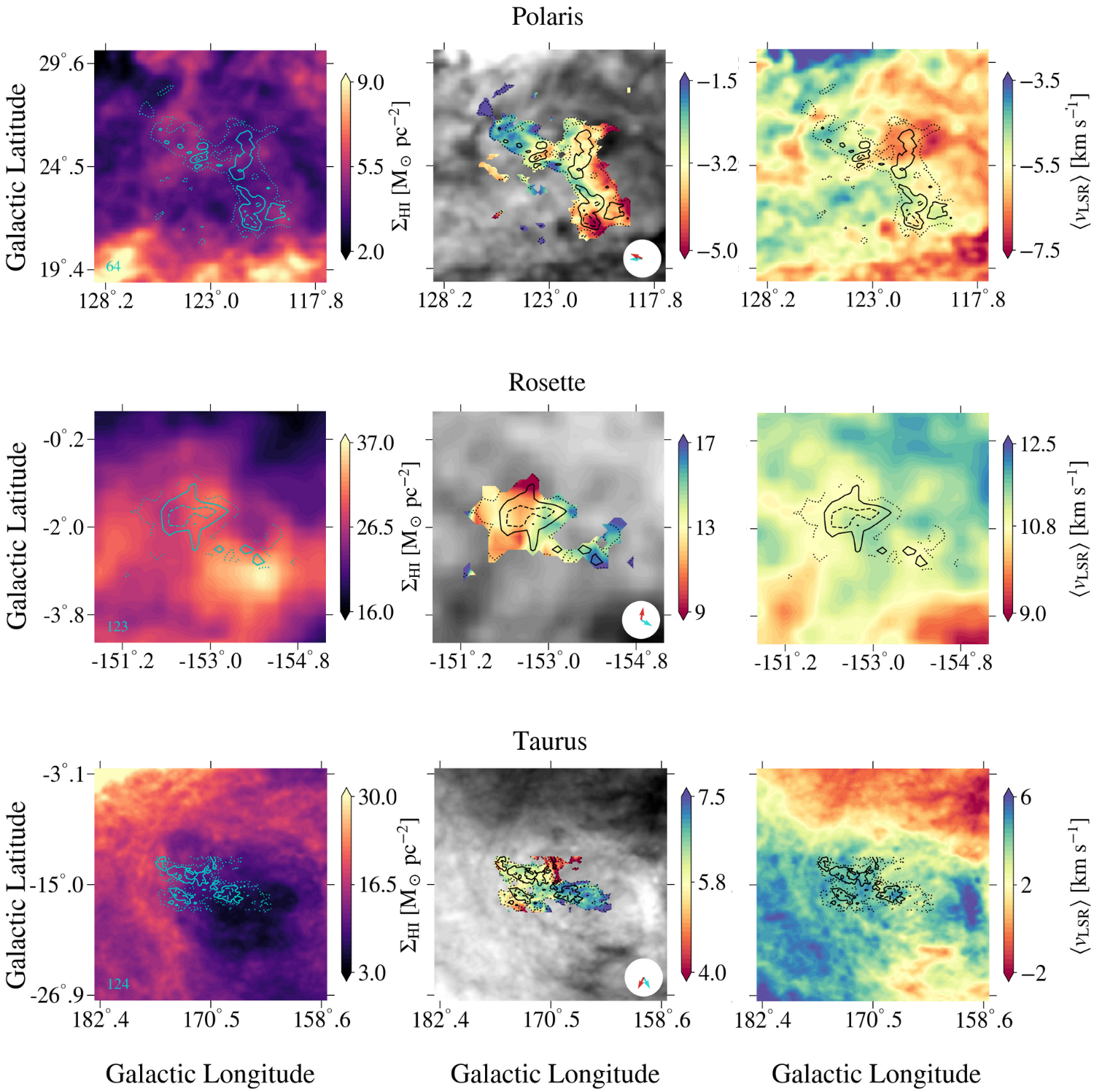}
\caption{Same as Figure \ref{fig:appndx_maps1}}
\label{fig:appndx_maps6}
\end{figure}
\newpage

\newpage
\section{Velocity Profiles}\label{sec:appndx_vvsr}

Velocity profiles used to determine the goodness of fit of a plane (Equation \ref{eq:v_plane}) for representing the velocity field maps of the molecular clouds and HI envelopes (Appendix \ref{sec:appndx_maps}). We use \texttt{scipy.optimize.curve\_fit} to calculate the least-squares fit, using the inverse second moment maps (available upon request) as weights.

\begin{figure}[h!]
\includegraphics[width=\linewidth]{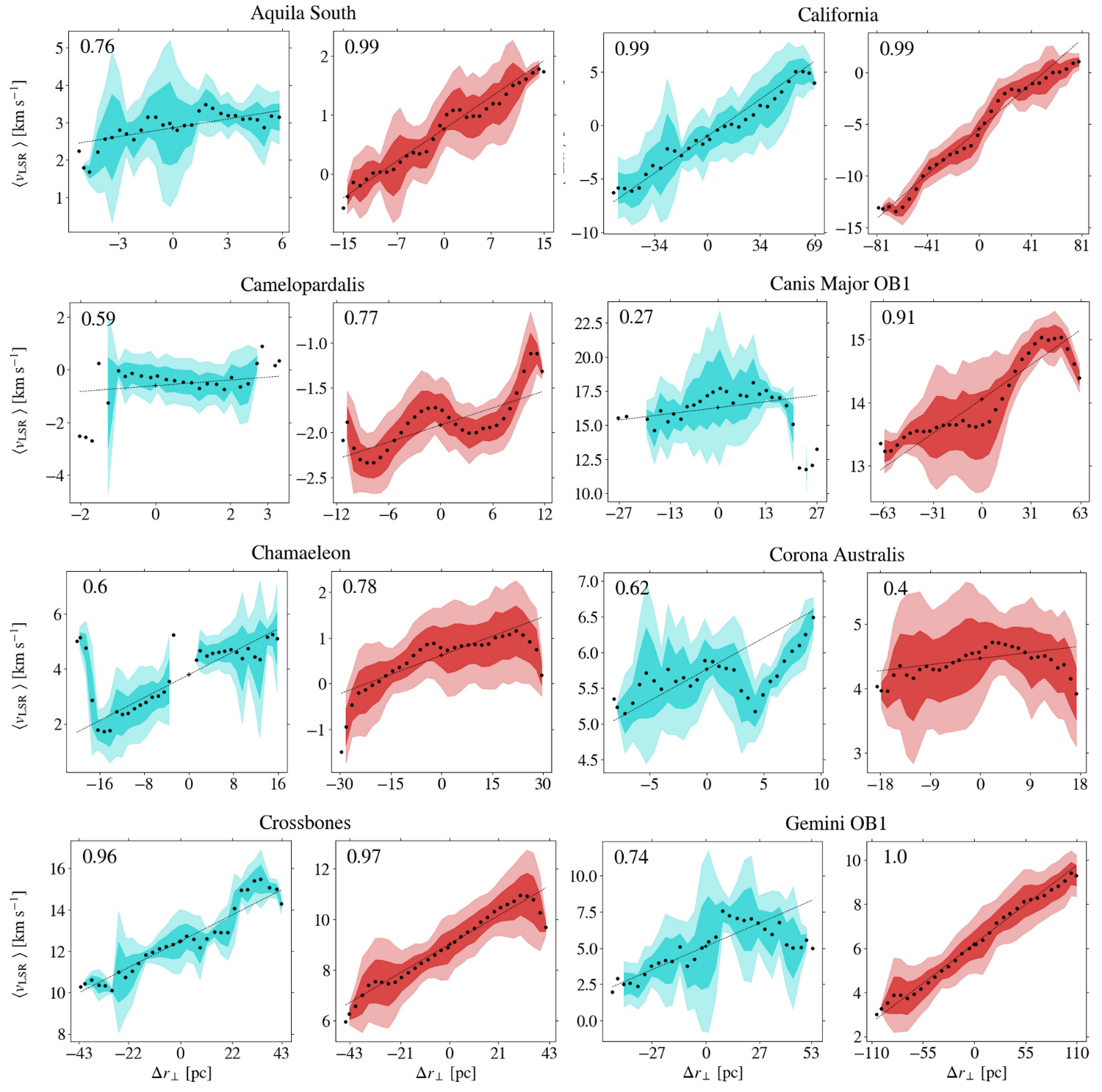}
\caption{Intensity-weighted velocity centroid, $\langle v_\mathrm{LSR} \rangle$, as a function of the perpendicular displacement, $r_\perp$, from the rotation axis of the \textbf{each} molecular cloud (\textit{left}) and its HI envelope (\textit{right}). The dashed lines indicate the planar model, and the shaded regions show the $\pm 1\sigma$ and $\pm 2\sigma$ scatter of the velocity field map at each radial bin. The Pearson correlation coefficient of the velocity profile is in the upper left corner. }
\label{fig:appndx_vvsr1}
\end{figure}

\begin{figure}[ht!]
\includegraphics[width=\linewidth]{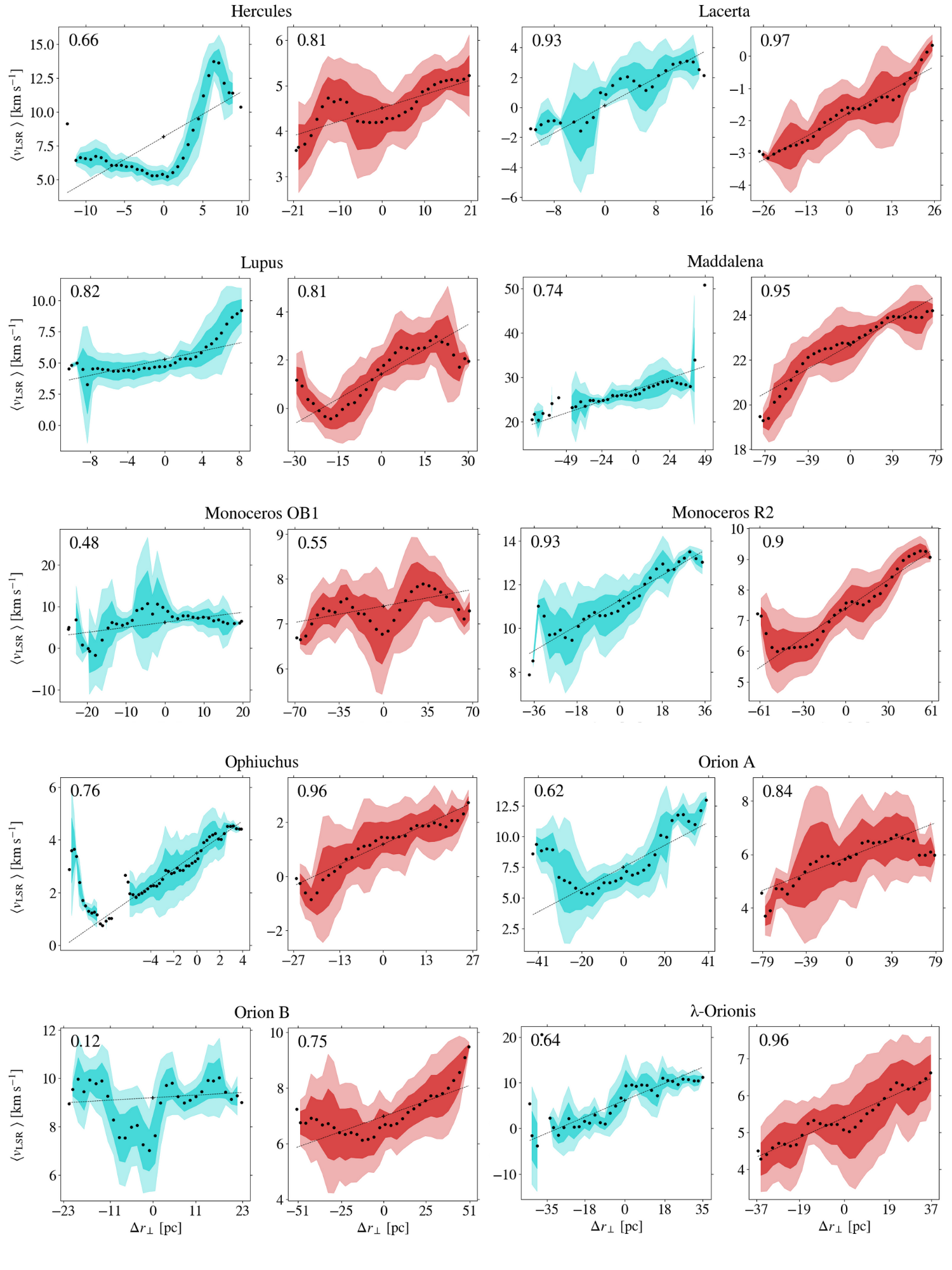}
\caption{Same as Figure \ref{fig:appndx_vvsr1}}
\label{fig:appndx_vvsr2}
\end{figure}

\begin{figure}[ht!]
\includegraphics[width=\linewidth]{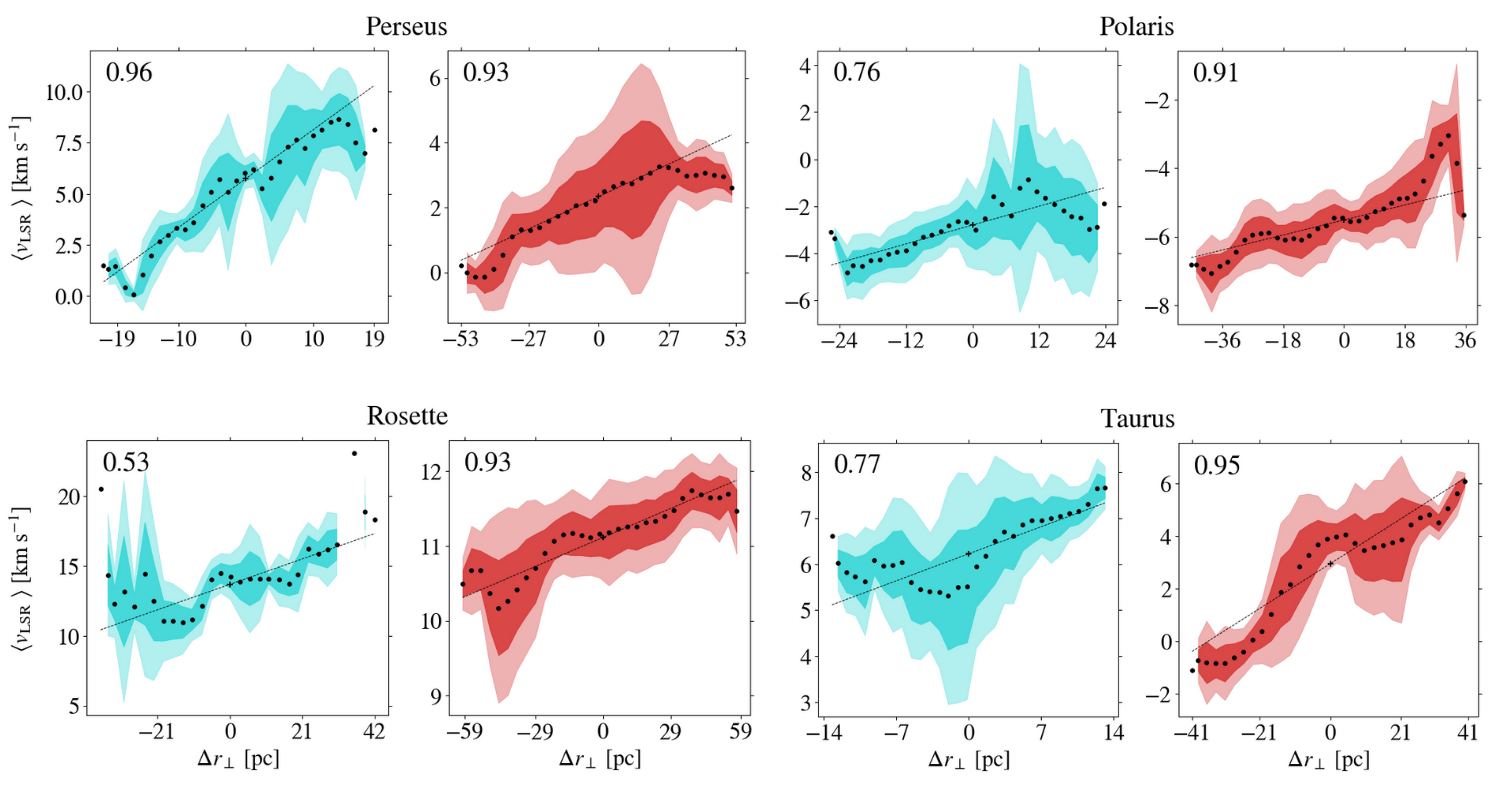}
\caption{Same as Figure \ref{fig:appndx_vvsr1}}
\label{fig:appndx_vvsr3}
\end{figure}

\end{document}